\documentclass[12pt]{article}
\usepackage{amsfonts}
\usepackage{amsmath}
\usepackage{amssymb}
\usepackage{array}
\usepackage{bigints}
\usepackage{booktabs}
\usepackage[nosort]{cite}
\usepackage{color}
\usepackage{dsfont}
\usepackage{float}
\usepackage{framed}
\usepackage{graphicx}
\usepackage{indentfirst}
\usepackage{mathrsfs}
\usepackage{multirow}
\usepackage{setspace}
\usepackage{subdepth}
\usepackage{subfig}
\usepackage{titlesec}
\usepackage[dotinlabels]{titletoc}
\usepackage{wrapfig}
\usepackage[all]{xy}
\usepackage{young}
\usepackage[vcentermath]{youngtab}
\usepackage{inputenc}
\usepackage{hyperref}
\hypersetup{colorlinks=true}
\hypersetup{linkcolor=black}
\hypersetup{citecolor=black}
\hypersetup{urlcolor=black}

\numberwithin{equation}{section}

\usepackage[left=2.5cm,right=2.5cm,top=2.5cm,bottom=3cm]{geometry}
\titleformat{\section}{\normalfont\bfseries}{\thesection.}{4pt}{}
\titlespacing{\section}{0pt}{20pt}{6pt}

\titleformat{\subsection}{\normalfont\itshape}{\thesubsection.}{4pt}{}
\titlespacing{\subsection}{0pt}{15pt}{6pt}

\titleformat{\subsubsection}{\normalfont}{\thesubsubsection.}{4pt}{}
\titlespacing{\subsubsection}{0pt}{15pt}{6pt}

\def\ie{\begin{equation}\begin{aligned}}
\def\fe{\end{aligned}\end{equation}}

\newcommand\VRule[1][\arrayrulewidth]{\vrule width #1}

\def\hat{\widehat}

\def\bar{\overline}

\def\half{{1 \over 2}}
\def\d{\partial}

\def\ep{\varepsilon}
\def\1{{\mathds 1}}

\DeclareMathOperator{\tr}{tr}
\DeclareMathOperator{\Tr}{\mathrm{Tr}}

\newcommand{\Z}{{\mathbb Z}}

\newcommand{\R}{{\mathbb R}}

\def\SL{{\mathscr L}}

\def\CA{{\mathcal A}}
\def\CB{{\mathcal B}}

\def\CD{{\mathcal D}}
\def\CE{{\mathcal E}}

\def\CI{{\mathcal I}}
\def\CJ{{\mathcal J}}

\def\CN{{\mathcal N}}
\def\CO{{\mathcal O}}

\def\CQ{{\mathcal Q}}

\def\CT{{\mathcal T}}

\def\CV{{\mathcal V}}

\DeclareFontShape{OT1}{cmr}{mx}{n}%
    {<->cmr10}{}
\newcommand{\mytitlefont}{\fontseries{mx}\selectfont}
\DeclareMathAlphabet{\titlemath}{OT1}{cmr}{mx}{n}

\begin{document}


\begin{titlepage}

\begin{center}

~\\[2cm]

{\fontsize{25pt}{0pt} \mytitlefont  ${\cal N}=(1,0)$ Anomaly Multiplet Relations \\[3pt] in Six Dimensions}

~\\[0.5cm]

Clay C\'{o}rdova,$^1$ Thomas T.~Dumitrescu,$^2$ and Kenneth Intriligator\,$^3$

~\\[0.1cm]

$^1$~{\it Kadanoff Center for Theoretical Physics \& Enrico Fermi Institute, University of Chicago}

$^2$\,{\it Mani L.\,Bhaumik Institute for Theoretical Physics, Department of Physics and Astronomy,}\\[-4pt]
       {\it University of California, Los Angeles}~\\[0.2cm] 

$^3$ {\it Department of Physics, University of California, San Diego}

~\\[0.8cm]

\end{center}

\noindent  We consider conformal and 't Hooft anomalies in six-dimensional~$\CN=(1,0)$ superconformal field theories, focusing on those conformal anomalies that determine the two- and three-point functions of conserved flavor and~$SU(2)_R$ currents, as well as stress tensors. By analyzing these correlators in superspace, we explain why the number of independent conformal anomalies is reduced in supersymmetric theories.  For instance, non-supersymmetric CFTs in six dimensions have three independent conformal~$c$-anomalies, which determine the stress-tensor two- and three-point functions, but in superconformal theories the three~$c$-anomalies are subject to a linear constraint. We also describe anomaly multiplet relations, which express the conformal anomalies of a superconformal theory in terms of its 't Hooft anomalies. Following earlier work on the conformal~$a$-anomaly, we argue for these relations by considering the supersymmetric dilaton effective action on the tensor branch of such a theory. We illustrate the utility of these anomaly multiplet relations by presenting exact results for conformal anomalies, and hence current and stress-tensor correlators, in several interacting examples. 

\vfill

\begin{flushleft}
December 2019
\end{flushleft}

\end{titlepage}


\tableofcontents

\section{Introduction}

\subsection{Conformal Anomalies}\label{sec:introconf}

The hallmark of conformal invariance in local quantum field theory is the existence of a traceless stress tensor $T_{\mu \nu}$.  However, when a conformal field theory (CFT) in even spacetime dimension $d$ is coupled to background fields, such as a gauge field sourcing a conserved current $J_{\mu},$ or a metric sourcing $T_{\mu \nu},$ quantum anomalies can lead to a non-vanishing trace.  In this case, the one-point function $\langle T_\mu ^\mu \rangle$ is a quantity of scaling dimension $d$ constructed from the background field strengths.  The structure of these expressions in the background fields is universal and can be fixed using symmetries, while the coefficients of the allowed terms are theory-dependent conformal anomalies.

Standard conformal anomalies can also be interpreted directly in terms of the underlying CFT, without background fields. Then~$T^{\mu}_{\mu}$ vanishes as an operator, and hence has trivial correlation functions at separated points, but it has non-vanishing contact terms dictated by the anomalies.  Such contact terms are ultimately related to stress-tensor correlators at separated points, and hence they provide meaningful and useful information about the CFT. 

In this paper we focus on conformal anomalies of CFTs in $d=6$ dimensions.  Coupling such a theory to a curved background metric leads to four possible anomalies, whose coefficients are conventionally called~$a$ and $c_{1}, c_{2}, c_{3}$ \cite{Deser:1976yx, Duff:1977ay, Fradkin:1983tg, Deser:1993yx},
\begin{equation}\label{6dWeyl}
\langle T^{\mu }_{\mu }\rangle \; \supset \; a E+ c_{1}I_{1}+c_{2}I_{2}+c_{3}I_{3}~.
\end{equation} 
Here~$E$ is the Euler density in six dimensions, which is given in terms of the Riemann curvature two-form~$R^{ab}$ as
\begin{equation}\label{Eis}
E \; \sim \; \varepsilon_{a_{1}a_{2}a_{3}a_{4}a_{5}a_{6}}R^{a_{1}a_{2}}R^{a_{3}a_{4}}R^{a_{5}a_{6}}~,
\end{equation}
while the $I_{i}$ are expressions constructed out of the Weyl curvature tensor (for precise expressions see e.g.~\cite{Deser:1993yx}),
\begin{equation}\label{Isare}
I_{1}\sim W_{\mu \phantom{\nu\rho}\nu}^{\phantom{\mu}\rho \sigma \phantom{D}}W_{\rho\phantom{EF}\sigma}^{\phantom{B}\chi \psi \phantom{C}}W_{\chi \phantom{AD}\psi}^{\phantom{E}\mu \nu \phantom{F}}~,\hspace{.15in}I_{2}\sim W_{\mu\nu\rho \sigma}W^{\rho \sigma \chi \psi}W_{\chi \psi}^{\phantom{EF}\mu\nu }~,\hspace{.15in}I_{3}\sim W_{\mu\nu\rho\sigma}\nabla^{2}W^{\mu\nu\rho \sigma} + \, \cdots~.
\end{equation}
In terms of correlation functions at separated points, the coefficient $c_3$ determines the two-point function of $T_{\mu \nu}$, while $c_{1}, c_{2}, c_{3}$ fix the three-point function of $T_{\mu \nu}$. The anomaly coefficient $a$ first arises in a four-point function of stress tensors. We fix a convenient convention for SCFTs in $d = 6$ such that a single, free~$\CN =(2,0)$ tensor multiplet has $a=c_{1}=c_{2}=c_{3}=1.$ Table \ref{knownW} lists the conformal anomalies of some six-dimensional CFTs in these conventions. 
\begin{table}[h]
\centering
\begin{tabular}{!{\VRule[1pt]}c!{\VRule[1pt]}c!{\VRule[1pt]}c!{\VRule[1pt]}c!{\VRule[1pt]} c!{\VRule[1pt]}}
\specialrule{1.2pt}{0pt}{0pt}
{\bf Theory} & $\bf a$ &  $\bf c_{1}$ & $\bf c_{2}$ & $\bf c_{3}$ \\
\specialrule{1.2pt}{0pt}{0pt}
\multirow{2}{*}{ Scalar}& \multirow{2}{*}{$\frac{1}{441}$} &  \multirow{2}{*}{$\frac{1}{180}$} & \multirow{2}{*}{$-\frac{1}{252}$} & \multirow{2}{*}{$\frac{1}{70}$} \\
 &  & & &  \\
 \hline
 \multirow{2}{*}{ Weyl Fermion}& \multirow{2}{*}{$\frac{191}{4410}$} &  \multirow{2}{*}{$\frac{4}{45}$} & \multirow{2}{*}{$\frac{4}{105}$} & \multirow{2}{*}{$\frac{1}{7}$} \\
 &  & & &  \\
 \hline
  \multirow{2}{*}{ Chiral Two-Form}& \multirow{2}{*}{$\frac{221}{245}$} &  \multirow{2}{*}{$\frac{143}{180}$} & \multirow{2}{*}{$\frac{1189}{1260}$} & \multirow{2}{*}{$\frac{9}{14}$} \\
 &  & & &  \\
 \hline
\multirow{2}{*}{(1,0) Hypermultiplet}& \multirow{2}{*}{$\frac{11}{210}$} &  \multirow{2}{*}{$\frac{1}{9}$} & \multirow{2}{*}{$\frac{1}{45}$} & \multirow{2}{*}{$\frac{1}{5}$} \\
 &  & & &  \\
\hline
\multirow{2}{*}{(1,0) Tensor multiplet}& \multirow{2}{*}{$\frac{199}{210}$} &  \multirow{2}{*}{$\frac{8}{9}$} & \multirow{2}{*}{$\frac{44}{45}$} & \multirow{2}{*}{$\frac{4}{5}$} \\
 &  & & &  \\
 \hline
 \multirow{2}{*}{$(2,0)$ Theory with algebra $\frak{g}$}& \multirow{2}{*}{$\frac{16}{7}h^\vee _{\frak{g}}d_{\frak{g}}+r_{\frak{g}}$} &  \multirow{2}{*}{$4h^\vee _{\frak{g}}d_{\frak{g}}+r_{\frak{g}} $} & \multirow{2}{*}{$4h^\vee _{\frak{g}}d_{\frak{g}}+r_{\frak{g}}$} & \multirow{2}{*}{$4h^\vee _{\frak{g}}d_{\frak{g}}+r_{\frak{g}}$}\\
 &  & & &  \\
\specialrule{1.2pt}{0pt}{0pt}
 \end{tabular}
\caption{Conformal anomalies of some~$d = 6$ CFTs.  The first three lines are the minimal free field theories in six dimensions. The two subsequent lines are the free $\CN=(1,0)$ SCFTs.  The final line summarizes the interacting $\CN=(2,0)$ SCFTs.  These are labelled by an ADE Lie algebra $\frak{g}$, and their conformal anomalies are determined in terms of the rank $r_{\frak{g}},$ the dimension $d_{\frak{g}}$, and the dual coxeter number $h^\vee _{\frak{g}}$ of $\frak{g}$.  The large-$N$ behavior of these conformal anomalies was established in \cite{Henningson:1998gx}. The exact formula for $a$ was computed in \cite{Cordova:2015vwa}. The exact formula for the $c$-anomalies of $\CN=(2,0)$ theories was conjectured in \cite{Beem:2014kka} and derived in \cite{Cordova:2016cmu}.  }
\label{knownW}
\end{table}

In CFTs with continuous flavor symmetries, there are additional conformal anomalies. To lighten notation, we focus on a single factor~$G_F$ of the full flavor symmetry; here~$G_F$ could be either~$U(1)$ or simple. In the presence of background gauge fields (with field strength~$F_{\mu\nu}$) that source the associated conserved flavor current~$J_\mu^F$, these flavor conformal anomalies take the following form,
\begin{equation}\label{Fconfanom}
\langle T_\mu ^\mu \rangle \supset  - \frac{\tau^F_{1}}{3} \Tr ({\cal D}_\mu F ^{\mu \lambda }{\cal D}^\nu F_{\nu  \lambda})-\frac{\tau^F_{2}}{4}W_{\mu \nu \rho \lambda}\Tr (F^{\mu \nu}F^{ \rho\lambda})+\frac{\rho^F}{3} \Tr (F^{\mu \nu}F_{\mu \lambda}F^\lambda{}_\nu) + \cdots~.
\end{equation}
Here we omitted several additional terms that are fixed by conformal symmetry; we also dropped total derivative terms. We conclude that every~$U(1)$ or simple factor $G_F$ of the flavor symmetry gives rise to three conformal anomaly coefficients: $\tau_{1}^{F}, \tau_{2}^{F}$, and~$\rho^{F}$. Note that the anomaly controlled by $\rho^F$, which is cubic in the field strength $F_{\mu\nu}$, is proportional to the structure constants $f^{abc}$ of~$G_F$ and hence vanishes for abelian flavor symmetries.

As we will review below, these conformal anomalies encode basic data about correlation functions of conserved currents in the CFT at separated points and in the absence of background fields. Specifically, $\tau_{1}^{F}$ determines the two-point function~$\langle J^FJ^F\rangle$, while~$\tau_{1}^{F}$, $\tau_{2}^{F}$ determine the three-point function $\langle TJ^{F}J^{F}\rangle$, and $\tau_{1}^{F}$, $\rho^{F}$ determine the three-point function  $\langle J^{F}J^{F}J^{F}\rangle$. We see that conformal anomalies completely determine all two- and three-point functions of flavor currents and the stress tensor. 

\subsection{Supersymmetry Constraints on Conformal Anomalies}

In general, non-supersymmetric CFTs, the three $c_i$ anomalies are independent. Below we will prove that $\CN =(1,0)$ SCFTs only have two  independent $c_i$ anomalies. In such theories, the three~$c_i$ coefficients are related by a universal, linear relation dictated by supersymmetry, 
\begin{equation}\label{susycrelation}
c_1=\half \left(c_2+c_3\right)~.
\end{equation}
As we will see, one consequence of this relation is that the two distinct unitary~$\CN=(1,0)$ free field theories, i.e.~the free hypermultiplet and the free tensor multiplet, span the space of stress-tensor supermultiplet three-point functions.

The relation~\eqref{susycrelation} was conjectured in~\cite{deBoer:2009pn}, where it was noted that it is satisfied by all free-field and holographic examples, and that it is analogous to a known relation obeyed by four-dimensional SCFTs in the context of conformal collider physics~\cite{Hofman:2008ar} (see below). A linear relation between the~$c_i$ anomalies is also suggested by the more recent observation~\cite{Butter:2016qkx} that minimal conformal supergravity in~$d = 6$ appears to only admit two independent candidate invariants that can supersymmetrize the~$I_i$ invariants in~\eqref{6dWeyl} and~\eqref{Isare}. 

In section \ref{SCFTcursec} we establish \eqref{susycrelation} via a superspace analysis. We show that the superspace three-point function of stress-tensor supermultiplets~$\CT(x_i, \theta_i)$ takes the general form
\begin{equation}\label{TTTform}
\langle \CT (x_1,\theta _1)\CT (x_2, \theta_2)\CT(x_3, \theta_3)\rangle \sim H_{\CT\CT\CT}(Z)=C_1^{\CT\CT\CT} \frac{1}{X^4}+C_2^{\CT\CT\CT}\frac{(X\Theta)^2}{X^8}+ C_3^{\CT\CT\CT}\frac{\Theta ^8}{X^8}~,
\end{equation}
which depends on three real constants $C_{1,2,3}^{\CT\CT\CT}$ that are linearly related to the~$c_i$ anomalies. (Here $Z=(X,\Theta)$ is a superspace variable that is constructed using the three superspace coordinates~$(x_i, \theta_i)$.)  We then argue that the conservation equation obeyed by the stress-tensor multiplet imposes one linear constraint on the coefficients $C_{1,2,3}^{\CT\CT\CT}$, which in turn leads to the relation~\eqref{susycrelation} among the~$c_i$.   

We also use superspace to analyze those supermultiplet three-point functions that can be obtained from~\eqref{TTTform} by replacing one, two, or all three stress-tensor multiplets by conserved flavor-current multiplets.  In every case supersymmetry imposes one additional linear relation on the coefficients of the corresponding non-supersymmetric correlators. It follows that some of the conformal anomalies related to flavor symmetries, which were defined in~\eqref{Fconfanom}, vanish in all~$\CN=(1,0)$ SCFTs,\footnote{~The fact that $\rho^F=0$ in SCFTs agrees with observations in~\cite{Manvelyan:2000ea} based on free-field reasoning and the AdS$_7$ supergravity duals of $\CN = (2,0)$ theories.}
\begin{equation}
\label{susytaufrel}
\tau^{F}_{2}= \rho^{F} =0~.
\end{equation}
To streamline the notation, we will often write
\begin{equation}\label{susytauone}
\tau_{1}^{F}\equiv \tau_{F}
\end{equation}
for the unique non-vanishing conformal anomaly coefficient associated with the flavor symmetry~$G_F$. The relations above imply that~$\tau_{F}$ governs both the~$\langle J^FJ^F\rangle$ two-point function, as well as the~$\langle T J^F J^F\rangle$ and~$\langle J^FJ^FJ^F\rangle$ three-point functions.   

In terms of conformal collider observables~\cite{Hofman:2008ar}, the fact that~$\tau _2^F=0$ implies that the energy flux $\langle {\cal E}(\widehat n)\rangle$ in a state created by the flavor current is independent of its polarization.  Similarly, the fact that  $\rho ^F=0$ implies that the charge flux $\langle {\cal Q}(\widehat n)\rangle$ in such a state is also independent of the polarization.  These properties of flavor currents in six-dimensional SCFTs are direct analogues of relations that hold for SCFTs in four dimensions~\cite{Hofman:2008ar}. 
 
Every~$\CN=(1,0)$ SCFT has an~$SU(2)_R$ symmetry. The associated conserved current~$J^R_\mu$ resides in the stress-tensor supermultiplet, and consequently the conformal anomaly coefficients~$\tau_{1,2}^R$ and~$\rho^R$ it gives rise to need not satisfy the relations~\eqref{susytaufrel} that hold for flavor currents. Adapting the arguments outlined above to this case, we find that the conformal anomaly coefficients associated with the~$SU(2)_R$ symmetry can be expressed using two linearly independent~$c_i$ anomaly coefficients (with the third one given by~\eqref{susycrelation}), 
\begin{equation}\label{susytaurrel}
 \tau_{1}^{R}=c_3~, \qquad \tau_{2}^{R}=\half \left(c_3-c_1\right)~,\qquad \rho^{R} =\frac{3}{2} \left(c_1-c_3\right)~.
\end{equation}

Let us consider the special case of~$\CN=(2,0)$ theories, which have an~$Sp(4)_R$ symmetry. When viewed as~$\CN=(1,0)$ theories, it is natural to focus on the maximal subgroup~$SU(2)_R \times SU(2)_F \subset Sp(4)_R$. Here~$SU(2)_R$ and~$SU(2)_F$ are $R$- and flavor symmetries that are distinguished by the choice of~$\CN=(1,0)$ subalgebra, but they are exchanged by a Weyl reflection inside the full~$Sp(4)_R$ symmetry. It follows that the vanishing conditions~\eqref{susytaufrel} that hold for the~$SU(2)_F$ symmetry must also apply to~$SU(2)_R \subset Sp(4)_R$, i.e.~$\tau_2^R = \rho^R = 0$.\footnote{~This agrees with a conjecture of~\cite{Manvelyan:2000ef} that was motivated by free-field and holographic reasoning.} Comparing with~\eqref{susytaurrel} (and using~\eqref{susycrelation}), we conclude that the~$c_i$ anomalies of all~$\CN=(2,0)$ theories necessarily coincide, in agreement with the explicit formulas in Table~\ref{knownW}.

\subsection{Anomaly Multiplets}\label{sec:introam}

So far we have discussed conformal anomalies (i.e.~'t Hooft anomalies for conformal symmetry), which arose from various conserved currents such as the stress tensor or a flavor current. These currents may themselves have more conventional 't Hooft anomalies, which lead to violations of current conservation in the presence of background fields. Such anomalies are conveniently summarized by an anomaly 8-form, from which the anomalous transformation of the effective action follows via the standard descent procedure (see e.g.~\cite{Cordova:2018cvg} for a detailed recent discussion with references).  For instance, 't Hooft anomalies for the~$SU(2)_R$ symmetry or diffeomorphisms are characterized by the following anomaly polynomial,\footnote{~Here $c_{i}(R)$ denotes the Chern class of degree $2i$ for the $SU(2)_R$ background gauge bundle, while $p_{i}(T)$ denotes the Pontryagin class of degree $4i$ for the tangent bundle.}
\begin{equation}\label{alphadelta}
\mathcal{I}_{8}\supset \frac{1}{4!}\left(\alpha c_{2}(R)^{2}+\beta c_{2}(R)p_{1}(T)+\gamma p_{1}(T)^{2}+\delta p_{2}(T) \right)~.
\end{equation}
The corresponding 't Hooft anomaly coefficients $\alpha, \beta, \gamma, \delta$ are universal and independent observables in any six-dimensional $\CN=(1,0)$ SCFT.  They are often exactly calculable, e.g.~via string constructions and anomaly inflow arguments~\cite{Harvey:1998bx, Ohmori:2014pca} or by analyzing RG flows~\cite{Ohmori:2014kda, Intriligator:2014eaa, Heckman:2015ola, WIP}.  We enumerate these anomaly coefficients for several~$\CN=(1,0)$ theories in Table \ref{known}. 

\smallskip

\begin{table}[h]
\centering
\begin{tabular}{!{\VRule[1pt]}c!{\VRule[1pt]}c!{\VRule[1pt]}c!{\VRule[1pt]}c!{\VRule[1pt]} c!{\VRule[1pt]}}
\specialrule{1.2pt}{0pt}{0pt}
{\bf Theory} & $\bf \alpha$ &  $\bf \beta$ & $\bf \gamma$ & $\bf \delta$ \\
\specialrule{1.2pt}{0pt}{0pt}
\multirow{2}{*}{ Hypermultiplet}& \multirow{2}{*}{$0$} &  \multirow{2}{*}{$0$} & \multirow{2}{*}{$7 \over 240$} & \multirow{2}{*}{$-{1 \over 60}$} \\
 &  & & &  \\
\hline
\multirow{2}{*}{(1,0) Tensor multiplet}& \multirow{2}{*}{$1$} &  \multirow{2}{*}{$\half$} & \multirow{2}{*}{$23 \over 240$} & \multirow{2}{*}{$-{29 \over 60}$} \\
 &  & & &  \\
 \hline
 \multirow{2}{*}{(1,0) Vector multiplet}& \multirow{2}{*}{$-1$} &  \multirow{2}{*}{$- \half $} & \multirow{2}{*}{$-{7 \over 240}$} & \multirow{2}{*}{$1 \over 60$}\\
 &  & & &  \\
\hline
 \multirow{2}{*}{$(2,0)$ Theory with algebra $\frak{g}$}& \multirow{2}{*}{$h^\vee _{\frak{g}}d_{\frak{g}}+r_{\frak{g}}$} &  \multirow{2}{*}{$\half r_{\frak{g}} $} & \multirow{2}{*}{$\frac{1}{8}r_{\frak{g}}$} & \multirow{2}{*}{$-\half r_{\frak{g}}$}\\
 &  & & &  \\
\specialrule{1.2pt}{0pt}{0pt}
 \end{tabular}
\caption{$SU(2)_R$ and diffeomorphism 't Hooft anomaly coefficients of some~$\CN=(1,0)$ theories. Note that the free vector multiplet is not conformal, but nevertheless enjoys an $SU(2)_R$ symmetry.  The $\CN=(2,0)$ anomalies for $\frak{g} = \frak{su}(N)$ were first computed in~\cite{Harvey:1998bx}. The general formulas for any ADE Lie algebra~$\frak{g}$ were conjectured in~\cite{Intriligator:2000eq} and verified in \cite{Yi:2001bz, Ohmori:2014kda,Cordova:2015vwa}. }
\label{known}
\end{table}

It is an important and general fact that supersymmetry relates conformal anomalies and 't Hooft anomalies. This means that the typically challenging and delicate conformal anomalies can be analyzed using the more accessible and robust 't Hooft anomalies. In the case of~$d = 4$ SCFTs, such relations where established in~\cite{Anselmi:1997am}  by analyzing the anomalous stress-tensor supermultiplet of the SCFT in the presence of background supergravity fields -- an object often referred to as the anomaly multiplet.  In~\cite{Cordova:2015fha} we derived an anomaly multiplet relation between the conformal anomaly $a$ and 't Hooft anomaly coefficients in~\eqref{alphadelta},
\begin{equation}\label{ais}
a=\frac{16}{7}(\alpha-\beta+\gamma)+\frac{6}{7}\delta~. 
\end{equation}
Rather than analyzing the anomalous stress-tensor supermultiplet in~$d = 6$ SCFTs, this relation was derived by studying the dilaton effective action on the tensor branch of the SCFT -- a technique we will also utilize in this paper.\footnote{~A direct analysis of anomalous stress-tensor multiplets is more challenging in~$d = 6$ than in~$d = 4$. For instance, non-conformal stress-tensor multiplets (of which the anomaly multiplet is a special case) and the associated supergravity theories have been thoroughly analyzed in~$d = 4$ (see e.g.~\cite{Komargodski:2010rb, Dumitrescu:2011iu} and references therein), while their~$d = 6$ counterparts are not nearly as well studied. Moreover, a direct investigation of $d = 6$ anomaly multiplets via anomalous supercurrents requires detailed knowledge of certain~$R^{3}$ supergravity invariants, which is technically rather ominous. Some recent progress in this direction appears in~\cite{Butter:2016qkx}.} A consequence of this derivation was a proof of the~$a$-theorem for RG flows from the SCFT onto its tensor branch.\footnote{~It was shown in~\cite{Louis:2015mka, Cordova:2016xhm,Cordova:2016emh} that the only supersymmetric RG flows in six dimensions that start at a SCFT fixed point are flows onto the moduli space of vacua of that SCFT.} The behavior of $a$ under Higgs branch RG flows was subsquently explored in~\cite{Heckman:2015axa}, and the anomaly multiplet relation \eqref{ais} has been verified holographically in \cite{Cremonesi:2015bld}. 

In this paper, we likewise establish anomaly multiplet relations for the $c_i$ conformal anomalies in~\eqref{6dWeyl}, by expressing them in terms of the 't Hooft anomaly coefficients~$\alpha, \beta, \gamma, \delta$ in~\eqref{alphadelta} via the following formulas,
\begin{equation}\label{csare}
c_{1}=4\alpha -\frac{14}{3}\beta+\frac{16}{3}\gamma+\frac{8}{3}\delta~,\hspace{.2in}c_{2}=4\alpha -\frac{10}{3}\beta+\frac{8}{3}\gamma+\frac{10}{3}\delta~,\hspace{.2in}c_{3}=4\alpha -6\beta+8\gamma+2\delta~.
\end{equation}
Note that these formulas are compatible with the universal linear relation~$c_1 = \half\left(c_2 + c_3\right)$ in~\eqref{susycrelation}, even though $\alpha, \beta, \gamma, \delta$ are independent. 

In theories with flavor symmetries, there are additional 't Hooft anomaly coefficients that are visible in the presence of background flavor gauge fields. As above, we consider a single abelian or simple factor~$G_F$ of the full flavor symmetry, and we focus on the mixed anomalies of~$G_F$ with the~$SU(2)_R$ symmetry or diffeomorphisms, 
\begin{equation}
 \CI _8 \; \supset \; \frac{1}{4!}\left(\alpha _{F^2R^2} \, c_2(F)c_2(R)+\alpha _{F^2T^2} \, c_2(F)p_1(T)\right)~.
\end{equation}
We will argue that the 't Hooft anomaly coefficients~$\alpha_{F^2R^2}$ and~$\alpha_{F^2 T^2}$ determine the conformal anomaly coefficient~$\tau_F$ in~\eqref{susytauone} as follows,
\begin{equation}\label{taualpha}
\tau_{F}=2\alpha _{F^2T^2}-2\alpha _{F^2R^2}~.
\end{equation}

The formulas~\eqref{csare} have already appeared in the recent literature~\cite{Yankielowicz:2017xkf, Beccaria:2017dmw}. There linear relations between the~$c_i$ conformal anomalies and the 't Hooft anomaly coefficients~$\alpha,\beta, \gamma, \delta$ were postulated, and the unknown coefficients in these relations were fixed by considering examples. This is complicated by the fact that each anomaly multiplet relation is specified by four coefficients, while there are only three independent classes of unitary SCFTs for which all pertinent anomalies are reliably known (the free hyper- and tensor multiplets, and the $\CN=(2,0)$ theories). To circumvent this problem, the authors of~\cite{Yankielowicz:2017xkf, Beccaria:2017dmw} considered a non-unitary but superconformal free field theory constructed from an abelian vector multiplet  with a higher-derivative kinetic term as a fourth data point. This theory had previously been shown to satisfy the anomaly multiplet relation~\eqref{ais} for the~$a$ conformal anomaly~\cite{Beccaria:2015uta}.

In section~\ref{sec:dilaton}, we will instead follow~\cite{Cordova:2015fha} and argue for the anomaly multiplet relations~\eqref{csare} and \eqref{taualpha} by studying conformal and 't Hooft anomaly matching along RG flows from an SCFT onto its tensor branch. As in~\cite{Cordova:2015fha}, our main tool will be the supersymmetric dilaton effective action on the tensor branch.

\subsection{Application to the Small $E_{8}$ Instanton SCFTs}

The anomaly multiplet relations \eqref{csare} and \eqref{taualpha} have a wide variety of applications.  Many~$\CN=(1,0)$ SCFTs in six dimensions have been constructed in string theory, starting with~\cite{Seiberg:1996vs, Seiberg:1996qx} and generalizations in \cite{Ganor:1996mu, Blum:1997mm, Brunner:1997gk, Brunner:1997gf, Hanany:1997gh}, recently culminating in a systematic analysis using F-theory~\cite{Heckman:2013pva, Heckman:2015bfa, Heckman:2018jxk}.  Such theories have also been explored using holography \cite{Apruzzi:2013yva, Gaiotto:2014lca, Passias:2016fkm}. None of these~$\CN=(1,0)$ SCFTs possess a known Lagrangian description. They are essentially isolated (because they do not admit any supersymmetry-preserving relevant or marginal deformations~\cite{Louis:2015mka, Cordova:2016xhm,Cordova:2016emh}) and strongly coupled (see for instance~\cite{Chang:2018xmx}).  

Nevertheless, as we illustrate in section \ref{sec:examples}, the conformal anomalies of these strongly-coupled SCFTs can often be determined using anomaly multiplet relations. For instance, in the~$\CN=(1,0)$ SCFT described by $N$ small $E_{8}$ instantons in string theory \cite{Ganor:1996mu} we can use~\eqref{taualpha} to compute the two-point function coefficient~$\tau_{E_8}$ of the $E_{8}$ flavor currents, 
\begin{equation}
\tau_{E_{8}}=24N^{2}+36N~.
\end{equation}
This formula has already appeared in~\cite{Chang:2017xmr}, where it was found to agree with bootstrap results.

\section{Current and Stress-Tensor Two- and Three-Point Functions}\label{sec:corrrev}

In this section we review results from~\cite{Osborn:1993cr,Erdmenger:1996yc} about two- and three-point functions of conserved flavor currents~$J_\mu^a$ associated with a global flavor symmetry $G_F$, and the stress-tensor~$T_{\mu \nu}$. We explain how these correlation functions are related to conformal anomaly coefficients, completing various discussions in the existing literature. In this section we consider general CFTs, without assuming supersymmetry. For each two- and three-point current correlator, we first present results for a general spacetime dimension~$d$ before specializing to~$d = 6$. As above, we assume that the flavor symmetry~$G_F$ is abelian or simple. Flavor Lie algebra indices will be denoted by~$a, b, c, \ldots$\,.  

Two-point functions are completely determined by conformal symmetry, up to an overall coefficient. For flavor currents and the stress tensor we have
\begin{equation}\label{nonsusyjjTT}
\langle J_\mu ^a (x) J_\nu ^b (0)\rangle = \frac{C_F\delta ^{ab}}{x^{2(d-1)}} I_{\mu \nu}(x), \qquad \langle T_{\mu \nu}(x)T_{\rho \sigma}(0)\rangle = \frac{C_T}{x^{2d}}I_{\mu \nu, \rho \sigma}(x)~,
\end{equation} 
where
\begin{equation}
I_{\mu \nu}(x)=\delta_{\mu\nu}-2\frac{x_{\mu}x_{\nu}}{x^{2}}~, \hspace{.5in}I_{\mu\nu, \rho \sigma}=\frac{1}{2}\left(I_{\mu\sigma}I_{\nu\rho}+I_{\mu\rho}I_{\nu\sigma}\right)-\frac{1}{d}\delta_{\mu\nu}\delta_{\rho\sigma}~.
\end{equation} 

In~$d = 6$ dimensions, the two-point function coefficient $C_T$ is proportional to the conformal anomaly coefficient $c_3$ in~\eqref{6dWeyl}.  To determine the proportionality constant, it suffices to compare them for free-field CFTs. For a theory with~$n_\phi$ free real scalar fields and~$n_\psi$ free fermion fields (with~$\dim(\psi)$ complex spinor components), it was shown in~\cite{Osborn:1993cr} that 
\begin{equation}\label{CTis}
C_T=n_\phi \frac{d}{d-1}\frac{1}{S_d^2}+n_\psi \frac{d}{2}\dim(\psi)\frac{1}{S_d^2} \quad \longrightarrow \quad \frac{6}{5\pi ^6}\left(n_\phi+10n_\psi\right)~.
\end{equation}
Here $S_d=2\pi ^{ d \over 2}/\Gamma ({d \over 2})$; in the last expression we have set $d=6$ and taken $n_\psi$ to be the number of chiral fermions, with $\dim (\psi)=4$ complex components.   We can now compare with our normalization for~$c_3$ in Table \ref{knownW} to conclude that 
\begin{equation}
c_{3}\equiv \frac{\pi ^6}{84}C_T~.
\end{equation}    

Similarly, the flavor-current two-point function coefficient~$C_F$ is proportional to the conformal anomaly coefficient~$\tau^F_1$ in~\eqref{Fconfanom}. We will chose a convention for the proportionality factor that is convenient for six-dimensional SCFTs. It was shown in~\cite{Osborn:1993cr} that in free scalar theories, with flavor current $J_\mu ^a=\phi t^a_\phi \partial _\mu \phi$ (here~$t^a_\phi$ is real and antisymmetric), and in free fermion theories, with flavor current~$J_{\mu}^{a}=\bar \psi t_\psi ^a \gamma _\mu \psi$ (here~$t^a_\psi$ is complex and antihermitian) the coefficient~$C_F$ is given by 
\begin{equation}\label{CFis}
C_F= { T_{\phi} \over 2}\frac{1}{(d-2)S_d^2}+{ T_{\psi} \over 2} \text{dim}(\psi)\frac{1}{S_d^2} \quad \longrightarrow \quad \frac{1}{8\pi ^6}\left(T_\phi +16T_\psi\right)~.
\end{equation}
The first expression is valid for general $d$, while we have set~$d = 6$ in the second expression. The coefficients depend on the quadratic indices for the representations of the bosons or fermions, $\tr (t_{\phi, \psi}^a t_{\phi , \psi}^b)=-\half T _{\phi , \psi}\delta ^{ab}$. We choose our normalization convention for~$\tau^F _1$ such that the $Sp(4)_R$ symmetry that acts on a free $\CN=(2,0)$ tensor multiplet has $\tau _1^{Sp(4)_R}= 1$. This is equivalent to the statement that the~$SU(2)_F$ flavor symmetry that acts on a single, free $\CN=(1,0)$ hypermultiplet has $\tau _1^{F}= 1$. Thinking of such a free hypermultiplet as a half-hyper that transforms as an~$SU(2)_F$ doublet, we conclude that it has~$T_\phi =2$, $T_\psi = \half$, so that
\begin{equation}\label{tau1norm}
\tau_{1}^{F}\equiv \frac{4\pi ^6}{5}C_F~. 
\end{equation}
 
For generic~$d$, it was shown in~\cite{Osborn:1993cr,Erdmenger:1996yc} that the flavor-current three-point function $\langle J_\mu^a (x) J^b_\nu (y) J^c_\lambda (z)\rangle $ is fully determined by conformal symmetry and conservation laws, up to two coefficients ${\cal A}_{FFF}$ and ${\cal B}_{FFF}$,\footnote{~In $d=4$ there is also a parity-odd structure proportional to $d^{abc}$, whose coefficient determines the cubic 't Hooft anomaly of the flavor symmetry.}  
\begin{equation}\label{JJJAB}
\langle J_\mu ^a(x)J_\nu ^b(y)J_\lambda ^c(z)\rangle= f^{abc}\frac{I_{\nu \sigma}(x-y)I_{\lambda \rho}(x-z)I_{\sigma \beta}(X)}{(x-y)^{d-2}(x-z)^{d-2}(y-z)^d} \, t_{\mu \beta \rho}(X)~,
\end{equation}
where $f^{abc}$ are the structure constants and
\begin{equation}\label{eq:abfffdef}
X_\mu = \frac{(x-y)_\mu}{(x-y)^2}-\frac{(x-z)_\mu}{(x-z)^2}~, \quad t_{\mu \nu \lambda}(X)={\cal A}_{FFF}\frac{X_\mu X_\nu X_\lambda}{X^2}+{\cal B}_{FFF}(X_\mu \delta _{\nu \lambda}+X_\nu \delta _{\mu \lambda}-X_\lambda \delta _{\mu \nu})~.
\end{equation}
A Ward identity implies that~\cite{Osborn:1993cr,Erdmenger:1996yc}
\begin{equation}
C_F=S_d\left(\frac{1}{d} {\cal A}_{FFF} +{\cal B}_{FFF}\right) \quad \longrightarrow \quad \pi ^3\left(\frac{{\cal A}_{FFF}}{6}+{\cal B}_{FFF}\right)~.
\end{equation}
Thus the~$\langle JJJ\rangle$ three-point function introduces one additional, theory-dependent constant beyond the~$\langle JJ\rangle$ two-point function coefficient~$C_F$.

The free-field values of the coefficients ${\cal A}_{FFF}$ and ${\cal B}_{FFF}$ were computed in~\cite{Osborn:1993cr},
\begin{equation}\label{ffab}
\begin{split}
& \mathcal{A}_{FFF}=\frac{d \, T_{\phi} }{4(d-2)S_{d}^{3}}\; \longrightarrow\;\frac{3}{8\pi ^9} T_\phi~, \\
& \mathcal{B}_{FFF}=\frac{T_{\phi}}{4(d-2)S_{d}^{3}}+\frac{T_{\psi} \dim(\psi)}{2S_{d}^{3}} \; \longrightarrow \; \frac{1}{\pi ^9}\left(\frac{T_\phi}{16}+2T_\psi\right)~.
\end{split}
\end{equation}
Instead of using~$\CA_{FFF}, \CB_{FFF}$ to parameterize the~$\langle JJJ \rangle$ three-point function, we can also express it in terms of free scalar or fermion correlators, 
\begin{equation}\label{nJJJ}
\langle J_\mu ^a(x)J_\nu ^b(y)J_\lambda ^c(z)\rangle=n^{FFF}_\phi \langle J_\mu ^a(x)J_\nu ^b(y)J_\lambda ^c(z)\rangle_\phi +n^{FFF}_\psi \langle J_\mu ^a(x)J_\nu ^b(y)J_\lambda ^c(z)\rangle_\psi~.
\end{equation}
The two coefficients $n^{FFF}_\phi$ and $n^{FFF}_\psi$ are linearly related to ${\cal A}_{FFF}$ and ${\cal B}_{FFF}$. The exact relation can be extracted from the free-field results summarized above. 

The stress-tensor three-point function $\langle T_{\mu \nu}(x)T_{\rho \sigma}(y)T_{\kappa \lambda}(z)\rangle$ is also determined by conformal symmetry and conservation laws up to three coefficients \cite{Osborn:1993cr, Erdmenger:1996yc}.  A Ward identity relates one linear combination of these three coefficients to the two-point function coefficient~$C _T$.  In~$d = 6$ dimensions, the three stress-tensor three-point function coefficients are linearly related to the three conformal anomalies $c_1, c_2, c_3$ in~\eqref{6dWeyl}.  We can span the three structures using the stress-tensor three-point functions of free fields: a free scalar $\phi$, a free Weyl fermion $\psi$, and a free, chiral two-form $B$ (with self-dual three-form field strength $H$), see~\cite{Bastianelli:1999ab} for details,
\begin{equation}\label{nTTT}
\langle TTT\rangle = n_{\phi}^{TTT}\langle TTT\rangle _\phi+n_{\psi} ^{TTT} \langle TTT\rangle _\psi +n_{B}^{TTT}\langle TTT\rangle _B~.
\end{equation}
In a free theory, the coefficients~$n_{\phi, \psi, B}^{TTT}$ coincide with the number~$n_\phi, n_\psi, n_B$ of free scalars, Weyl fermions, or chiral two-forms, but in an interacting CFT they are defined by~\eqref{nTTT}. In general, these coefficients are constrained by unitarity and conformal collider inequalities (see section~\ref{subsconfcol}). Comparing to the known free-field conformal anomalies (see~\cite{Fradkin:1982kf,Bastianelli:1999ab} and Table \ref{knownW}) we conclude that
\begin{equation}
\begin{split}\label{csfree}
& c_1= \frac{n^{TTT}_\phi +16n^{TTT}_\psi +143 n^{TTT}_B}{180}~, \\
& c_2= \frac{-5n^{TTT}_\phi +48n^{TTT}_\psi +1189 n^{TTT}_B}{1260}~,\\
&  c_3=\frac{n^{TTT}_\phi +10 n^{TTT}_\psi +45 n^{TTT}_B}{70}~.
\end{split}
\end{equation}

We now consider the three-point function~$\langle T JJ\rangle$ of one stress tensor and two flavor currents, which was also analyzed in~\cite{Osborn:1993cr,Erdmenger:1996yc} and shown to be determined by two coefficients. One linear combination of these coefficients is proportional to the~$\langle J J\rangle$ two-point function coefficient~$C_F$ (or equivalently to $\tau _1^F$, see~\eqref{tau1norm}) thanks to a Ward identity, while the remaining independent structure constant can be thought of as the OPE coefficient of the stress tensor in the fusion of two flavor currents. 

In the notation of \cite{Osborn:1993cr,Erdmenger:1996yc} (see for instance the discussion around equation (3.14) of~\cite{Osborn:1993cr}), the two coefficients that determine the~$\langle TJJ\rangle$ three-point function are called~$c_{TFF}$ and $e_{TFF}$. Their free-field values can be found in equation~(5.10) of \cite{Osborn:1993cr}; setting~$d \rightarrow 6$ and~$\text{dim}\left(\psi\right) = 4$ in these formulas, we find that 
\begin{equation}\label{cefree}
\pi ^9c_{TFF}=\frac{3}{5}T_\phi +12 T_\psi, \qquad \pi ^9e_{TFF}=\frac{3}{20}T_\phi~.
\end{equation}
It follows that the two independent structures in the $\langle TJJ\rangle$ three-point function are spanned by free field theories in which the flavor current only arises from charged scalars or fermions, 
\begin{equation}\label{njjT}
\langle TJJ\rangle = n^{TFF}_\phi \langle TJJ\rangle _\phi + n^{TFF}_\psi \langle TJJ\rangle _\psi~. 
\end{equation}

As reviewed in section \ref{sec:introconf}, a~$U(1)$ or simple flavor symmetry~$G_F$ gives rise to the conformal anomalies~\eqref{Fconfanom} in the presence of suitable background flavor and gravity fields,
\begin{equation}\label{Fconfanomii}
\langle T_\mu ^\mu \rangle \supset  - \frac{\tau^F_{1}}{3} \Tr ({\cal D}_\mu F ^{\mu \lambda }{\cal D}^\nu F_{\nu  \lambda})-\frac{\tau^F_{2}}{4}W_{\mu \nu \rho \lambda}\Tr (F^{\mu \nu}F^{ \rho\lambda})+\frac{\rho^F}{3} \Tr (F^{\mu \nu}F_{\mu \lambda}F^\lambda{}_\nu) + \cdots~.
\end{equation}
The anomaly coefficients $\tau^F _{1,2}$ and~$\rho^F$ are theory dependent. In the absence of background fields, each conformal anomaly in~\eqref{Fconfanomii} represents a contact term associated with a particular three-point function at separated points. To derive these relations, one can for instance follow~\cite{Freedman:1991tk,Osborn:1993cr} and work in position space using the method of differential regularization; alternatively, one can work in momentum space. 

It follows that the conformal anomalies $\tau^F _1$ and $\tau^F _2$ in~\eqref{Fconfanomii} must be related in a universal, linear way to the three-point function coefficients~$c_{TFF}$ and $e_{TFF}$ in~\eqref{cefree}, or equivalently to the coefficients~$n^{TFF}_{\phi,\psi}$ in~\eqref{njjT}. Similarly, the conformal anomalies~$\tau_1^F$ and~$\rho^F$ must have a universal, linear relation to the three-point function coefficients $\CA _{FFF}$ and $\CB _{FFF}$ in~\eqref{eq:abfffdef}. We can determine these universal linear relations by comparing with various free-field examples that have been worked out in the literature, see for instance~\cite{Fradkin:1982kf,Manvelyan:2000ef,Osborn:2015rna,Huang:2018hho}. Converting to our conventions, we find that free field theories satisfy 
 \begin{equation}\label{Fconfanomff}
 \tau^F_2=\frac{1}{45}\left(T_\phi -4T_\psi\right), \qquad \rho^F = -\frac{1}{30}\left(T_\phi -4T_\psi\right)~.
 \end{equation}
By comparing this to the free-field formulas~\eqref{ffab} and~\eqref{cefree} above, we conclude that 
\begin{equation}
\tau^F_2=\frac{\pi ^9}{45}\left(8e_{TFF}-\frac{1}{3}c_{TFF}\right)~,  \qquad \rho^F = \frac{\pi ^9}{15}\left({\cal B}_{FFF}-\frac{3}{2}{\cal A}_{FFF}\right)~.
\end{equation}
Together with~\eqref{tau1norm}, this completes the relations between two- and three-point function coefficients of currents and stress tensors, and the conformal anomalies associated with flavor symmetries in~\eqref{Fconfanomii}. We summarize these relations here,
\begin{equation}\label{fetcrelns}
 \tau^F _1=\frac{4\pi ^6}{5} C_F~, \quad \tau^F _2=\frac{\pi ^9}{45}\left(8e_{TFF}-\frac{1}{3}c_{TFF}\right)~,  \qquad \rho^F = \frac{\pi ^9}{15}\left({\cal B}_{FFF}-\frac{3}{2}{\cal A}_{FFF}\right)~.
 \end{equation}

\section{Current and Stress-Tensor Supercorrelators in Six-Dimensional SCFTs}\label{SCFTcursec}

In this section we examine the constraints of supersymmetry on two- and three-point functions of flavor currents and stress tensors in six-dimensional SCFTs.  We will show that the conformal anomalies of all~$\CN=(1,0)$ SCFTs satisfy the following universal relations,
\begin{equation}\label{eq:secgoal}
c_1=\half (c_2+c_3)~, \qquad \tau_{2}^F=\rho ^F=0~.
\end{equation}
Here~$G_F$ can be any~$U(1)$ or simple flavor symmetry. We also show that the conformal anomalies associated with the~$SU(2)_R$ symmetry can be expressed in terms of the~$c_i$ conformal anomalies as follows,
\begin{equation}\label{susytaurrelii}
 \tau_{1}^{R}=c_3, \qquad \tau_{2}^{R}=\half \left(c_3-c_1\right)~,\qquad \rho^{R} =\frac{3}{2} \left(c_1-c_3\right)~.
\end{equation}
Finally we briefly discuss conformal collider bounds on these anomaly coefficients.

\subsection{Free SCFTs}\label{ss:free}

We begin by considering a theory of $n_H$ free hypermultiplets and $n_T$ free tensor multiplets. As we will see below, some relations uncovered in this simple free-field context continue to hold for general $\CN =(1,0)$ SCFTs. In fact, the input from free field theories will be used in the general proof of these relations below. 

Using~\eqref{csfree} with $n^{TTT}_\phi = 4n_H+n_T$, $n_\psi^{TTT} = n_T+n_H$, and $n_B^{TTT}=n_T$, we find that
\begin{equation}\label{c123nhnt}
c_1=\frac{n_H+8n_T}{9}~, \qquad c_2=\frac{n_H+44n_T}{45}~, \qquad c_3=\frac{n_H+4n_T}{5}~.
\end{equation}
Note that these formulas are consistent with~\eqref{eq:secgoal}, and that a free~$\CN=(2,0)$ tensor multiplet with~$n_{H}=n_{T}=1$ indeed satisfies~$c_1 = c_2 = c_3 =1$.

In supersymmetric theories we distinguish between flavor symmetries, which commute with the supercharges, and~$R$-symmetries, which do not:
\begin{itemize}
\item {\it Flavor Symmetries:} Only hypermultiplets can carry flavor charge, with $T^F_\phi = 4 T^F_\psi = 4T^F_H$, where $T^F_H$ is the quadratic index of the hypermultiplet flavor representation. Thus, the free-field expression for the flavor-current two-point function coefficients is 
\begin{equation}\label{susytau1}
\tau _1^F=2T_H^F~.
\end{equation}
Note that a single free hypermultiplet has an~$SU(2)_F$ flavor symmetry with~$T^F_H=\half$ and hence~$\tau^F_1 =1$. This is easy to see by reformulating the theory as a half-hyper in the doublet representation of~$SU(2)_F$. 

The free-field expression for the $\langle J^FJ^FJ^F\rangle$ flavor-current three-point function is obtained from~\eqref{ffab} by setting~$T^F_\phi = 4T^F_H$ and $T^F_\psi =T^F_H$, so that
\begin{equation}\label{abfree}
{\cal A}_{FFF}=\frac{3}{2\pi ^9} T^F_H~, \qquad {\cal B}_{FFF}=\frac{3}{2}{\cal A}_{FFF}=\frac{9}{4\pi ^9}T^F_H~.
\end{equation}
For the $\langle T J^F J^F\rangle$ three-point function, it follows from~\eqref{cefree} that
\begin{equation}
c_{TFF}=\frac{72}{5\pi ^9}T^F_H~, \qquad e_{TFF}=\frac{1}{24}c_{TFF}=\frac{3}{5\pi ^9}T^F_H~.
\end{equation}
Substituting into the expressions~\eqref{fetcrelns} for the flavor conformal anomalies gives 
\begin{equation}\label{ttrhofree}
\tau _2^F=\rho ^F=0~.
\end{equation}

\item {\it $SU(2)_R$ Symmetry:} The scalars in each hypermultiplet transform as a complex~$SU(2)_R$ doublet, so that $T_\phi^R=2n_H$. By contrast, the fermions in every tensor multiplet transform as half-doublets of $SU(2)_R$, so $T_\psi ^R=\half n_T$. It follows that the free-field expression for the~$SU(2)_R$ current two-point function coefficient is given by
\begin{equation}\label{susytau1bis}
\tau _1^R=\frac{1}{5}\left(n_H+4n_T\right)=c_3~.
\end{equation}
The free-field expressions for the~$\langle J^R J^R J^R\rangle$ and~$\langle T J^R J^R\rangle$ three-point function coefficitents take the following form,
\begin{equation}\label{eq:freeabce}
\begin{split}
& {\cal A}_{RRR}=\frac{3}{4\pi ^9}n_H~, \qquad {\cal B}_{RRR}=\frac{1}{8\pi ^9}(n_H+8n_T)~,\\
& c_{TRR}=\frac{6}{5\pi ^9}(n_H+5n_T)~, \qquad e_{TRR}=\frac{3}{10\pi ^9}n_H~.
\end{split}
\end{equation}
Substituting into~\eqref{fetcrelns} and comparing with~\eqref{c123nhnt}, we find that
\begin{equation}\label{eq:freet2rrr}
\tau _2^R=\frac{2}{45}\left(n_H-n_T\right) = \half \left(c_3-c_1\right)~, \quad \rho ^R=\frac{1}{15}\left(n_T-n_H\right)=\frac{1}{4}\left(c_1-c_3\right) ~.
\end{equation}

\end{itemize}

\subsection{Supercorrelators for Conserved Supermultiplets: Overview}\label{ss:conssupov}

We will apply the superspace formalism of~\cite{Park:1998nra} to the two- and three-point functions of operators in conserved flavor-current and stess-tensor supermultiplets. Analogous considerations for four-dimensional SCFTs can be found in~\cite{Osborn:1998qu,Park:1999pd,Kuzenko:1999pi}. Here we will give a brief survey of the results that will follow from this analysis (see section~\ref{ssec:scorr} below for details): 

\begin{itemize}
\item The two-point functions of all operators in the stress-tensor supermultiplet, including $T_{\mu \nu}$ itself and the~$SU(2)_R$ current~$J^{R}_\mu$, are determined by the two-point functions of the bottom component (i.e.~the superconformal primary); all of them can be expressed in terms of the conformal anomaly~$c_3$. Likewise, the two-point functions of all operators in a flavor-current supermultiplet are determined by the two-point functions of its bottom component; all of them can be expressed in terms of the conformal anomaly~$\tau _1^F$. The two-point function of the flavor-current supermultiplet with the stress-tensor supermultiplet vanishes.  The fact that two-point functions of supermultiplets are determined by those of their bottom components reflects the absence of nilpotent invariants for two-point functions. Such invariants first appear at the level of three-point functions~\cite{Park:1998nra,Osborn:1998qu,Park:1999pd,Kuzenko:1999pi}.

\item All non-zero three-point functions of operators in the supermultiplet of a flavor-current $J^a_\mu$ are entirely determined by the Lie algebra structure constants $f^{abc}$ and the two-point function coefficient $\tau _1^F$.  This differs from the non-supersymmetric case, where ${\cal B}_{FFF}/{\cal A}_{FFF}$, or equivalently the conformal anomaly $\rho^F$ in~\eqref{Fconfanom}, are independent quantities that appear in three-point functions.  Superconformal symmetry thus fixes these quantities, and to determine their values it suffices to compare to the free-field formulas~\eqref{abfree} and~\eqref{ttrhofree}. This implies that any~$\CN=(1,0)$ SCFT in six dimensions satisfies 
\begin{equation}\label{JJJABratio}
\frac{{\cal B}_{FFF}}{{\cal A}_{FFF}}\bigg| _\text{SCFT}=\frac{3}{2}~, \qquad \rho^F\big|_{\text{SCFT}}=0~.
\end{equation}

\item All three-point functions involving one stress-tensor supermultiplet operator and two flavor-current supermultiplet operators, such as $\langle T J^F J^F \rangle$ itself, are completely determined by $\tau _1^F$.  This implies that the ratio $e_{TFF}/c_{TFF}$, and thus the conformal anomaly coefficient $\tau _2^F$, are determined by supersymmetry.  Again, their values can then be computed from the free-field case, so that all six-dimensional~$\CN=(1,0)$ SCFTs satisfy
\begin{equation}\label{ecratio}
\frac{e_{TFF}}{c_{TFF}}\bigg| _\text{SCFT}=\frac{1}{24}  \qquad\text{hence}\qquad \tau _2^F\big|_\text {SCFT}=0~.
\end{equation}

\item  All non-zero three-point functions of operators in the stress-tensor supermultiplet are completely determined by two coefficients, one of which is related to the two-point function coefficient~$c_3$ by a Ward identity.  In particular, this proves that the three conformal anomalies $c_1$, $c_2$, $c_3$ necessarily satisfy a linear relation in any~six-dimensional SCFT, as originally conjectured in~\cite{deBoer:2009pn}. As before, the coefficients in this linear relation are fixed by the free-field formulas~\eqref{c123nhnt},
\begin{equation}\label{creltext}
c_2=\half \left(c_1+c_3\right)~.
\end{equation}

Since the~$SU(2)_R$ current~$J^R$ also resides in the stress-tensor supermultiplet, it follows that its three-point functions are also linear combinations of $c_1$ and $c_3$; the  coefficients are determined by the free-field formulas~\eqref{eq:freeabce} and~\eqref{eq:freet2rrr}. It follows that all~$\CN=(1,0)$ SCFTs satisfy 
 \begin{equation}
\pi ^9{\cal A}_{RRR}=\frac{3}{4}\left(10c_3-9c_1\right)~, \qquad \pi ^9  {\cal B}_{RRR}=\frac{9}{8}c_1~, \qquad \tau _2^R=\frac{1}{2}\left(c_3-c_1\right)~. 
\end{equation}
 This is compatible with the Ward identity relating the~$SU(2)_R$ current three- and two-point functions: $\pi ^3(\frac{1}{6}{\cal A}_{RRR}+{\cal B}_{RRR})=C_R=5c_3/4\pi ^6$, which fits~\eqref{tau1norm} with $\tau _1^R=c_3$. 
 
Likewise, the coefficients $e_{TRR}$ and $c_{TRR}$ that determine the three-point functions of one stress tensor and two~$SU(2)_R$ currents are universal linear combinations of $c_1$ and $c_3$ that can be fixed by free-field reasoning (see again~\eqref{eq:freeabce} and~\eqref{eq:freet2rrr}),   
\begin{equation}\label{TRR}
\pi ^9 c_{TRR}=\frac{27}{10}c_1+\frac{9}{2} c_3~, \qquad \pi ^9 e_{TRR}=3c_3-\frac{27}{10}
c_1~,\qquad \rho ^R=\frac{1}{4}(c_1-c_3)~. 
\end{equation}

 \end{itemize}

The main goal of the superspace analysis below is to cut down the number of independent structures in the two- and three-point functions of flavor-current and stress-tensor supermultiplets by imposing the constraints of superconformal symmetry. Once the number of these structures is sufficiently small, their coefficients can be determined by free-field reasoning. 

The constraints of conformal symmetry on two- and three-point functions are standard: two-point functions are always determined by one overall coefficient. By contrast, three-point functions are described by finitely many tensor structures, whose exact number can depend on the Lorentz representations of the operators participating in the three-point functions. Both of these results follow from the fact that the bosonic conformal generators can be used to bring the spacetime coordinates of the operators appearing in these correlators to standard form (see for instance~\cite{Kravchuk:2016qvl} and references therein for a recent discussion).

In superspace, the~$Q$ and~$S$ supercharges can be used to set the Grassmann coordinates of two operators to prescribed values. Thus all two-point functions of operators residing in a supermultiplet can be expressed in terms of the two-point function of its bottom component. However, at the level of three-point functions, superconformal symmetry alone does not simplify the dependence of the supercorrelator on the third Grassmann coordinate. 

The upshot is that superspace three-point functions generally depend on a non-trivial Grassman variable~$\Theta$ that can be constructed from the superspace coordinates of the three supermultiplets inside the correlator. The coefficient functions that appear in a~$\Theta$-expansion of such a supercorrelator represent the three-point functions of the individual component operators. In the absence of additional constraints, these functions are all independent, and hence the three-point functions of~$Q$-descendant operators inside a superconformal multiplet are not in general determined by the three-point function of the superconformal primary.

Additional constraints do arise if some of the supermultiplets in the correlator are short. The shortening condition implies that all correlators involving null states of the multiplet must vanish. In some cases, such constraints are sufficient to determine the three-point functions of all~$Q$-descendants in terms of the three-point function of the superconformal primaries (see e.g.~\cite{Kuzenko:1999pi,Fortin:2011nq,Goldberger:2012xb} for a related discussion in four dimensions).

As we will discuss below,  the superspace three-point function $\langle \CT (z_1)\CT (z_2)\CT (z_3)\rangle$ of the stress-tensor multiplet can be expressed in terms of a homogeneous function $H^{\CT\CT\CT}(X, \Theta)$ (here the bosonic variable~$X$ is the superpartner of~$\Theta$), which is completely determined up to three independent coefficients ${\cal C}_{1,2,3}^{\CT\CT\CT}$. We then impose the shortening condition on $\CT$, which amounts to setting a level-three null state and its descendants to zero. As we will see, this imposes one linear relation on the coefficients ${\cal C}_{1,2,3}^{\CT\CT\CT}$, so that the $\langle \CT (z_1)\CT (z_2)\CT (z_3)\rangle$ supermultiplet three-point function is in fact completely determined by two independent constants. Once this has been established, free-field reasoning is sufficient to deduce all facts about the stress-tensor supermultiplet three-point function that were summarized above.  

When analyzing the three-point supercorrelator~$\langle \CJ ^{(i_1j_1)}(z_1)\CJ ^{(i_2j_2)}(z_2)\CJ ^{(i_3j_3)}(z_3)\rangle$ of three flavor current multiplets, as well as supercorrelator $\langle \CJ ^{(i_1j_1)}(z_1)\CJ ^{(i_2j_2)}(z_2)\CT(z_3)\rangle$ of two flavor-current multiplets and one stress-tensor multiplet, we must impose the shortening condition satisfied by $\CJ ^{(ij)}$ in addition to that of $\CT$. This completely determines both correlators up to one overall coefficient, which in turn is related to the flavor-flavor two-point function coefficient $\tau _1^F$ by a Ward identity.

\subsection{Supercorrelators for Conserved Supermultiplets: Details}\label{ssec:scorr}

$\CN =(1,0)$ superspace in~$d = 6$ involves spacetime positions $x^{\mu}\in  [0,1,0]_{-1}^{(0)}$,\footnote{~We follow the conventions of~\cite{Cordova:2016xhm,Cordova:2016emh} and write the quantum numbers of various objects as~$[j_1, j_2, j_3]_\Delta^{(R)}$. Here $[j_1, j_2, j_3]$ denotes the Lorentz representation in $d=6$ using $SU(4)\sim SO(6)$ Dynkin labels. Thus $[1,0,0]$ and $[0,0,1]$ are chiral spinors, while $[0,1,0]$ is the vector representation of $SO(6)$.  Meanwhile, $R$ is the Dynkin label characterizing the~$SU(2)_R$ representation (i.e.~$R$ is always an integer and the representation has dimension~$R + 1$) and~$\Delta$ is the conformal scaling dimension.} which we will often write as antisymmetric bispinors~$x^{[\alpha\beta]} \sim \ep^{\alpha\beta\gamma\delta} x_{[\gamma\delta]}$ (see below), and symplectic-Majorana Grassmann coordinates $\theta ^\alpha _i\in [0,0,1]^{(1)}_{-1/2}$, where $\alpha =1,\ldots, 4$ is a chiral spinor index and~$i = 1,2$ an~$SU(2)_R$ doublet index. The latter can be raised and lowered using~$\ep^{ij}, \ep_{ij}$. The supercharges and supercovariant derivatives act via the following differential operators, 
\begin{equation}\label{QandD}
\CQ_\alpha^i = {\d \over \d \theta_i^\alpha} + i  \theta^{\beta i} \d_{\beta\alpha}~, \qquad  \CD_\alpha^i = {\d \over \d \theta_i^\alpha} -i \theta^{\beta i} \d_{\beta\alpha}~, 
\end{equation}
which satisfy $\left\{\CQ_\alpha^i, \CQ_\beta^j\right\} =  2i\ep^{ij} \d_{\alpha\beta}~$, $\left\{\CD_\alpha^i, \CD_\beta^j\right\} = -2i\ep^{ij} \d_{\alpha\beta}$, and $\left\{\CD_\alpha^i, \CQ_\beta^j\right\} = 0~.$

In the notation of~\cite{Cordova:2016xhm,Cordova:2016emh}, the stress-tensor of $\CN =(1,0)$ SCFTs resides in a~$B_3[0,0,0]_4^{(0)}$ superconformal multiplet with $40_B+40_F$ bosonic and fermionic component operators,
\begin{equation}\label{6dTops}
\begin{split}
& \CT \in [0,0,0]_4^{(0)}~, \qquad \psi _{\CT , \, \alpha}^i \in [1,0,0]^{(1)}_{4.5}~, \qquad J_{\mu}^{R (ij)}\in [0,1,0]_5^{(2)}~,  \\  
& C_{\CT, (\alpha \beta)} \in [2,0,0]_5^{(0)}~, \qquad  S_{ \mu\alpha}^i\in [1,1,0]^{(1)}_{5.5}~, \qquad T_{\mu \nu}\in [0,2,0]^{(0)}_6~.
\end{split}
\end{equation}
The  null states of the multiplet, which must be set to zero, first occur at level three,
\begin{equation}
\ep ^{\alpha \beta \gamma \delta}Q^i_\alpha Q^j_\beta Q^k_\gamma  \CT = (\CV ^\delta)^{(ijk)}\in [0,0,1]^{(3)}_{5.5}\quad \longrightarrow \quad 0~.\label{6dTnull}
\end{equation}
This enforces the conservation equations~$\partial ^\mu J_\mu^{R (ij)}=\partial ^\mu S_{\mu \alpha}^i=\partial ^\mu T_{\mu \nu}=0$ and ensures that~$S^i_{ \mu\alpha}$ and $T_{\mu \nu}$ are suitably traceless. In superspace, the shortening condition~\eqref{6dTnull}, and the resulting field content \eqref{6dTops}, can be expressed as follows, \begin{equation}
\ep ^{\alpha\beta\gamma\delta} \CD ^{i}_\alpha \CD^j_\beta \CD^{k}_\gamma \CT=0~, \qquad \CT(x, \theta)=\CT (x)+\psi^i_{\CT, \alpha}(x)\theta ^{\alpha}_i+J_{[\alpha \beta]}^{R (ij)} (x)(\theta \theta )^{[\alpha \beta]}_{(ij)} + \cdots~.\label{Tsuperspace}
\end{equation}

A conserved~$\CN=(1,0)$ flavor current resides in a~$D_1[0,0,0]_4^{(2)}$ multiplet~\cite{Cordova:2016xhm,Cordova:2016emh}, which contains $8_B+8_F$ operators, 
\begin{equation}
\CJ^{a (ij)}\in [0,0,0]_4^{(2)}~, \qquad J^{a i}_\alpha \in [1,0,0]^{(1)}_{4.5}~, \qquad J^a_{\mu} \in [0,1,0]_5^{(0)}~. 
\end{equation}
Here~$a$ is an adjoint flavor index. The conservation law for the flavor current~$J_\mu^a \sim J^a_{[\alpha\beta]}$ is encoded by the level-one null state $[1,0,0]_{4.5}^{(3)}  \rightarrow 0$ and its descendants. In superspace,
\begin{equation}\label{Jnull}
\CD ^{(i|}_\alpha \CJ ^{a | jk)}(x,\theta)=0~, \qquad \CJ ^{a(ij)}(x,\theta)=\CJ ^{a (ij)} (x)+J^{a ( i}_\alpha (x)\theta ^{j)\alpha}+\theta ^{i\alpha}\theta ^{j\beta}J^a_{[\alpha\beta]}(x) + \cdots~.
\end{equation}

Two-point functions can be written in terms of coordinates that are invariant under superspace translations~\cite{Park:1998nra}, 
\begin{equation}\label{chiis}
\chi ^{\alpha\beta}_{12}\equiv x_{1-}^{\alpha \beta}-x_{2+}^{\alpha\beta}+2i\theta _{2i}^\alpha \theta _1^{i\beta}~, \qquad \theta _{12, i}^\alpha \equiv \theta _{1,i}^\alpha -\theta _{2,i}^\beta~,
\end{equation}
where
\begin{equation}
x^{\alpha\beta}_\pm\equiv x^{\alpha \beta}\pm i \theta ^\alpha _i \theta ^{\beta i} = -x_{\mp}^{\beta \alpha}~.
\end{equation}
If the bottom component of the supermultiplet transforms in a non-trivial $SU(2)_R$ representation, the~$R$-symmetry indices of the supercorrelation function are accounted for by the quantities~\cite{Park:1998nra} (see also~\cite{Kuzenko:1999pi,Liendo:2015ofa} for a similar analysis in the context of~$d = 4$,~$\CN=2$ SCFTs)
\begin{equation}\label{uis}
{u_i}^j (z_{12})={\delta _i}^j -4i\theta _{12,i}\chi _{12}\theta _{12}^j (\det \chi _{12})^{-1/2}~.
\end{equation}

The superspace two-point function of the stress-tensor supermultiplet~$\CT(z)$  is completely determined up to an overall coefficient~$c_T\sim C_T$, giving the superspace generalization of~\eqref{nonsusyjjTT},
\begin{equation}
\langle \CT (z_1)\CT (z_2)\rangle = \frac{c_{T}}{(\det \chi _{12})^2}~.    \label{TTcorrelator}
\end{equation}
The two-point functions of all operators in the $\CT$-multiplet follow upon expanding in  $\theta _1, \theta_2$. In particular, the $J_{\mu}^{R(ij)}$ and $T_{\mu \nu}$ two-point functions are obtained from~\eqref{TTcorrelator} by extracting the $\theta _1^2\theta _2^2$ and the $\theta _1^4\theta _2^4$ terms, respectively.  Comparing with the general expressions~\eqref{nonsusyjjTT} for current and stress-tensor two-point functions, it follows that $c_T\sim \tau _1^{R}\sim c_3$.  The proportionality constants can be fixed by comparing to a free $\CN =(1,0)$ hyper- or tensor multiplet.  For a free tensor multiplet, the bottom component is $\CT | \sim \varphi ^2$, where $\varphi$ is the free scalar in the tensor multiplet;  for the free hypermultiplet, the bottom component is $\CT| \sim h_{i'i}h ^{i'i}$, where $i=1,2$ is an $SU(2)_R$ doublet index, $i'$ is an $SU(2)_F$ flavor doublet index, and the hypermultiplet scalars~$h_{i'i}$ obey a reality condition~$(h_{i' i})^\dagger \sim h^{i'i}$~\cite{Howe:1983fr}.\footnote{~See~\cite{Cordova:2018acb} for a related recent discussion in the context of~$d = 4$, $\CN=2$ theories.}

Similarly, the two-point functions of flavor-current supermultiplets $\CJ ^{a (ij)}$ are completely determined up to an overall coefficient $c_J$, 
\begin{equation}\label{supspff}
\langle \CJ ^a _{(i_ij_1)}(z_1) \CJ ^{b (i_2j_2) }(z_2) \rangle = c_J \, \delta ^{ab} \, \frac{u_{i_1}{}^{i_2}(z_{12})u_{j_1}{}^{j_2}(z_{12})+u_{i_1}{}^{j_2}(z_{12}){u_{j_1}}^{i_2}(z_{12})}{(\det \chi _{12})^2}~.
\end{equation}
The factors of ${u_i}^j(z_{12})$ (defined in~\eqref{uis}) account for the~$SU(2)_R$ representation of~$\CJ^{a(ij)}$. The actual flavor-current two-point function arises from the~$\theta _1^2\theta _2^2$ component of~\eqref{supspff}.  Finally, the two-point function of a flavor-current and a stress-tensor multiplet must vanish, 
\begin{equation}
\langle \CT (z_1) \CJ ^{a (ij)}(z_2)\rangle =0~.
\end{equation}
For the bottom components, this follows from~$SU(2)_R$ symmetry, while superconformal symmetry ensures that the same is true for all other operators in these multiplets. 

We now turn to the three-point functions involving the~$\CT$ and~$\CJ^{a(ij)}$ multiplets. As in~\cite{Osborn:1993cr,Park:1998nra}, they can be expressed in terms of a superspace variable $Z\equiv (X^\mu, \Theta ^{\alpha i})\in \R^{6|8}$ that is formed from the three superspace coordinates $z_{1,2,3}$ as follows,
\begin{equation}
X_{[\alpha\beta]}=\left(\chi _{13}^{-1}\chi _{12}\chi _{32}^{-1}\right)_{[\alpha\beta]}~, \qquad \Theta ^{i}_\alpha=i\left(\chi _{13}^{-1}\theta ^i_{13}-\chi ^{-1}_{23}\theta ^i_{23}\right)_\alpha~.
\end{equation}
The matrix~$\chi_{12}^{\alpha\beta}$ was defined in~\eqref{chiis} (the extension to other pairs of points is obvious), while~$\left(\chi^{-1}_{12}\right)_{\alpha\beta}$ is its inverse matrix. Note that $X_{[\alpha\beta]}\in [0,1,0]^{(0)}_1$ and $\Theta ^{i}_\alpha \in [1,0,0]^{(1)}_{1/2}$ have different scaling dimensions than the superspace coordinates~$x^\mu \in [0,1,0]^{(0)}_{-1}$ and $\theta^{\alpha i} \in [0,0,1]^{(1)}_{-1/2} $ discussed above~\eqref{QandD}; moreover~$\Theta^i_\alpha$ and $\theta^{\alpha i}$ are spinors of opposite chirality. 

The three-point supercorrelator of any Lorentz- and~$SU(2)_R$ singlet operators takes the form
\begin{equation}\label{threeOs}
\langle \CO _1(z_1)\CO _2(z_2)\CO _3(z_2)\rangle = \frac{H(Z_3)}{({\rm det} \chi _{13})^{\Delta _1/2}({\rm det} \chi _{23})^{\Delta _2/2}}~, 
\end{equation}
where $H(\lambda X, \lambda ^{1/2}\Theta)=\lambda ^{\Delta _3-\Delta _1-\Delta _2}H(X, \Theta)$.  The function $H$ can always be written as follows, 
\begin{equation}\label{threepointH}
H(X,\Theta)=\frac{1}{(X^2)^{\half (\Delta _1+\Delta _2-\Delta _3)}}\left( C^{\CO _1\CO _2\CO _3}_1+C^{\CO_1\CO _2\CO _3}_2 \frac{(X\Theta ^2)^2}{X^4}+C^{\CO _1\CO _2\CO _3}_3\frac{\Theta ^8}{X^4} \right).
\end{equation}
Here $(\Theta^2)_{(\alpha \beta)}\equiv \Theta ^{i}_\alpha\Theta ^{j}_\beta\ep _{ij}\in [2,0,0]^{(0)}$ is the only $SU(2)_R$ invariant that can be formed from the $\Theta ^{i}_\alpha$. It transforms in the same Lorentz representation as a self-dual three-form~$H^{[IJK]}$, with $I,J,K=1,\dots, 6$ and $*H=H$.  The only independent Lorentz scalar that can be constructed purely from $(\Theta^2)_{(\alpha\beta)}$ is $\Theta ^8\equiv \ep^{\alpha\beta\gamma\delta}\ep^{\alpha '\beta'\gamma'\delta'}(\Theta^2)_{\alpha \alpha'}(\Theta^2)_{\beta \beta'}(\Theta^2)_{\gamma \gamma'}(\Theta^2)_{\delta \delta'}$. (In terms of $H^{IJK}$, viewed as a trivalent vertex, this index contraction is a tetrahedron.) All remaining Lorentz and~$SU(2)_R$ invariants are built using the combination ${(X\Theta^2)^\alpha}_\beta  = X^{[\alpha\gamma]}(\Theta ^2)_{\gamma\beta }$, which is in the adjoint representation $[1,0,1]$ of the Lorentz group; it can also be written as $(X\Theta ^2)^{[IJ]}=H^{IJK}X_K$.  The quadratic Casimir invariant of this Lorentz adjoint, $\Tr (X\Theta^2)^2 \equiv (X\Theta ^2)^2$ appears as the middle term in~\eqref{threepointH}.
Note that the Lorentz adjoint $(X\Theta^2)$ does not have a cubic Casimir invariant, since $\Tr (X\Theta ^2)^3=0$. Finally, the quartic Casimir~$\Tr (X\Theta ^2)^4$ is not independent of the invariants that were already introduced above. 
 
If we expand the three-point function~\eqref{threepointH} in components, the coefficient $C^{\CO _1\CO _2\CO _3}_1$ determines the three-point function of the superconformal primaries.  The coefficients $C^{\CO_1\CO _2\CO _3}_{2,3}$ are associated with three-point functions of descendant operators. For long multiplets, the coefficients~$C^{\CO _1\CO _2\CO _3}_{1,2,3}$ are independent, but for short multiplets they are satisfy additional constraints that follow from the requirement that the null-state multiplet vanish in superspace. 

Let us apply the general reasoning above to the the three-point function of stress-tensor multiplets. Setting $\CO _i(z_i)=\CT (z_i)$ and $\Delta _i=4$ in~\eqref{threepointH}, we obtain
\begin{align}\label{HforTTT}
\langle \CT _1(z_1)\CT _2(z_2)\CT _3(z_2)\rangle &= \frac{H_{\CT\CT\CT}(X, \Theta)}{({\rm det} \chi _{13})^{2}({\rm det} \chi _{23})^{2}}~,\\ 
H_{\CT\CT\CT}(X, \Theta)&=C^{\CT\CT\CT}_1\frac{1}{X^4}+C^{\CT\CT\CT}_2 \frac{(X \Theta ^2)^2}{X^8}+C^{\CT\CT\CT}_3 \frac{\Theta ^8}{X^{8}}~. \nonumber
\end{align}
We must now impose the shortening condition~\eqref{6dTnull}, as well as exchange symmetry under $z_1\leftrightarrow z_2\leftrightarrow z_3$, since the three supermultiplets are identical.  In the present context, exchanging~$z_1\leftrightarrow z_2$ takes $Z\to -Z$ and does not constrain the coefficients in~\eqref{HforTTT}.\footnote{~By contrast, exchange symmetry does constrain analogous correlators in four-dimensional SCFTs, see for instance~\cite{Osborn:1998qu,Liendo:2015ofa}.} However, the shortening condition~\eqref{6dTnull} can be shown to lead to the following constraint on the function~$H_{\CT\CT\CT}$ that appears in~\eqref{HforTTT}, 
\begin{equation}\label{Hconstraint}
\ep_{\alpha\beta\gamma\delta}\hat \CD ^{\alpha}_i\hat \CD_{j}^\beta \hat \CD_{k}^\gamma H _{\CT\CT\CT}=0~, \qquad  \hat \CD_i^\alpha \equiv {\partial \over \partial \Theta_\alpha^i}+i \Theta _{i \beta}{\partial \over \partial X_{\beta \alpha}}~.
\end{equation}
To analyze this constraint, we count monomials in~$X^{[\alpha\beta]}$ and~$\Theta^{i}_\alpha$ that share its $[1,0,0]^{(3)}$ Lorentz and $R$-symmetry quantum numbers and could therefore appear as terms in~\eqref{Hconstraint}.  Naively, there are two such monomials: $[(X\Theta)^3] ^{(ijk)}_\alpha \equiv \ep_{\alpha\alpha_1\alpha_2\alpha_3} X^{\alpha _1\beta_1}X^{\alpha _2\beta _2}X^{\alpha _3\beta_3}\Theta ^{i}_{\beta _1}\Theta ^{j}_{\beta _2}\Theta ^{k}_{\beta _3}$; and $[\Theta ^5]^{(ijk)}_\alpha$, the unique contraction of five~$\Theta$'s with~$[1,0,0]$ Lorentz quantum numbers and~$R = 3$. However, $[(X\Theta )^3]^{(ijk)}_\alpha=0$ vanishes identically. This can be seen by 
noting that the $\Theta ^3$ part has to be in the $[0,0,1]^{(3)}$, which can give a $[1,0,0]$ Lorentz representation by tensoring with a~$[2,0,0]$. However, the latter is a three-form (i.e.~it is a completely antisymmetric tensor product of three~$[0,1,0]$ vectors), while~$X^3$ is a completely symmetric product of three vectors. Therefore their contraction~$(X\Theta)^3 \sim X^3 \Theta^3$ vanishes identically. 

It follows that evaluating the differential operator on the left-hand side of~\eqref{Hconstraint} on the function~$H_{\CT\CT\CT}$ in~\eqref{HforTTT} can (schematically) only give terms of the form 
\begin{equation}\label{nullterms}
\ep_{\alpha\beta\gamma\delta}\hat \CD ^{\alpha}_i\hat \CD_{j}^\beta \hat \CD_{k}^\gamma H_{\CT\CT\CT} (X, \Theta) \sim [\Theta ^5]_{\delta(ijk)}~. 
\end{equation}
Since the operator~$\hat \CD^\alpha_i$ changes the number of $\Theta$'s by $\pm 1$, the~$\hat \CD^3$ operator on the left-hand side can only produce the number of~$\Theta$'s on the right-hand side by acting on the terms proportional to~$C^{\CT\CT\CT}_2$ and $C^{\CT\CT\CT}_3$ in~\eqref{HforTTT}. It follows that~$C^{\CT\CT\CT}_1$ is unconstrained, and that there is one linear constraint on~$C^{\CT\CT\CT}_2$ and~$C^{\CT\CT\CT}_3$. This shows that all three-point correlators of operators in the stress-tensor supermutiplet can be expressed in terms of at most two linearly independent coefficients.   Moreover, there is a Ward identity that relates one linear combination of these two coefficients to the two-point function coefficient~$c_T\sim c_3$. As outlined in section~\ref{ss:conssupov}, it follows that the~$c_i$ conformal anomalies necessarily satisfy a linear relation in all~$\CN=(1,0)$ SCFTs. The precise form of this relation can be determined by examining free hyper- and tensor multiplets, as discussed around~\eqref{creltext}.

We can also use superspace to analyze three-point correlators that involve a flavor-current multiplet~\eqref{Jnull}.  This requires writing down general combinations of $u$'s, $\chi$'s, $X$, and $\Theta$ with the correct symmetries and quantum numbers, and then imposing the shortening conditions $(\CD ^3)^{(ijk)\alpha }\CT =0$ and $\CD _\alpha ^{(i}\CJ ^{jk)}=0$. The condition on the flavor current is particularly constraining, because the analogue of~\eqref{nullterms} now involves multiple independent structures on the right-hand side. The condition that all of these vanish separately determines the relative coefficients of all terms in the~$\Theta$-expansion. 

The upshot is that all three-point functions involving a flavor-current multiplet are fully determined up to an overall coefficient. Moreover, Ward identities relate this overall coefficient to a two-point function coefficient~$c_F \sim \tau_1^F$. The results are 
\begin{equation}\label{JTT}
\langle \CJ ^{a(i_ij_1)}(z_1) \CT (z_2) \CT (z_3)  \rangle =0~,
\end{equation}
and
\begin{equation}\label{JJJ}
\langle \CJ ^{a(i_1j_1) }(z_1)\CJ ^{b(i_2 j_2)}(z_2)\CJ ^{c(i_3,j_3)}(z_3)\rangle =c_F f^{abc} \langle \CJ^{(i_1j_1)}(z_1)\CJ^{(i_2j_2)}(z_2)\CJ^{(i_3j_3)}(z_3)\rangle _\text{canonical}~,
\end{equation} 
and 
\begin{equation}\label{JJT}
 \langle \CJ ^{a (i_ij_1)}(z_1) \CJ ^{b (i_2j_2)}(z_2) \CT (z_3)  \rangle = c_F\delta ^{ab}  \langle \CJ ^{(i_ij_1)}(z_1) \CJ ^{(i_2j_2)}(z_2) \CT (z_3)  \rangle_\text{canonical}~.
\end{equation}
Here the canonical correlators on the right-hand side are specific, completely determined functions on superspace, which are fixed by exchange symmetries and null-state conditions.  This six-dimensional analysis is closely analogous to the four-dimensional analysis in~\cite{Kuzenko:1999pi}.  For this reason, we do not spell out all the details here. Once it has been established that the correlators in~\eqref{JJJ} and~\eqref{JJT} only depend on the theory under consideration via the two-point function coefficient~$c_F \sim~\tau_2^1$ and the flavor Lie algebra structure constants $f^{abc}$, we can use free-field reasoning as in sections~\ref{ss:free} and \ref{ss:conssupov} to derive all other results quoted there.

\subsection{Conformal Collider Inequalities}\label{subsconfcol}

The average null energy condition~\cite{Faulkner:2016mzt, Hartman:2016lgu} places unitarity bounds on the three-point function coefficients discussed above. These bounds are conveniently derived using the conformal collider setup of \cite{Hofman:2008ar, deBoer:2009pn, Buchel:2009sk}. (See~\cite{Hofman:2016awc, Chowdhury:2017vel, Cordova:2017dhq, Cordova:2017zej} for some related recent work.) As described in~\cite{deBoer:2009pn,Osborn:2015rna}, the six-dimensional version of the collider bounds of~\cite{Hofman:2008ar}  can be written as
\begin{align}\label{6dccs}
1-\frac{1}{5}t_2-\frac{2}{35}t_4&\geq 0~,\nonumber\\
1-\frac{1}{5}t_2-\frac{2}{35}t_4+\half t_2&\geq 0~,\\
1-\frac{1}{5}t_2-\frac{2}{35}t_4+\frac{4}{5}( t_2 +t_4)&\geq 0~, \nonumber
\end{align}
where $t_{2,4}$ are linear combinations of the  $c_i$ conformal anomalies given in~\cite{deBoer:2009pn}.   In terms of the parameterization in~\eqref{nTTT}, these inequalities take the simple form
\begin{equation}
n_\phi^{TTT}\geq 0~, \qquad n_\psi^{TTT}\geq 0~, \qquad n_B^{TTT}\geq 0~.
\end{equation}

It was pointed out in~\cite{Hofman:2008ar} that $t_4=0$ for~$d = 4$ SCFTs.  By analogy, it was conjectured in~\cite{deBoer:2009pn} that supersymmetry should impose the linear relation~$t_4 = 0$ on the $c_i$ conformal anomalies of~$d = 6$ SCFTs, and it was verified that this is indeed the case in free-field examples.  In our conventions, the condition~$t_4 = 0$ precisely coincides with the relation~\eqref{creltext} derived above. The remaining inequalities in~\eqref{6dccs} then reduce to to~\cite{deBoer:2009pn},
\begin{equation}\label{c1c3ratio}
-\frac{5}{3} \leq t_2\leq 5 \qquad \Longleftrightarrow \qquad \frac{5}{9}\leq \frac{c_1}{c_3}\leq \frac{31}{27}~.
\end{equation}
The lower bound is saturated by a free hypermultiplet.  In fact, the upper bound in~\eqref{c1c3ratio} can be strengthened to~${10 \over 9} < {31 \over 27}$ in SCFTs, and this stronger upper bound is saturated by a free tensor multiplet. Note that unitarity of the two-point function implies~$c_3 > 0$, so that~\eqref{c1c3ratio} gives~$c_1>0$. Similarly, using~\eqref{creltext} gives $(c_2/c_3)=2(c_1/c_3)-1$ so that~\eqref{c1c3ratio} implies
\begin{equation}
\frac{1}{9}\leq \frac{c_2}{c_3} \leq \frac{35}{27}~,
\end{equation}
and hence $c_2>0$. 

To derive stronger constraints on the $c_{i}$ in the case of SCFTs we follow~\cite{Hofman:2008ar} and apply the conformal collider constraints to states created by the $R$-current. Let us first review the non-supersymmetric conformal collider bounds on states created by a conserved flavor current~$J^F_\mu$ in~$d$ dimensions. As reviewed below~\eqref{csfree}, the~$\langle T J^F J^F\rangle$ correlator is parametrized by two coefficients, $c_{TFF}$ and~$e_{TFF}$, one linear combination of which is fixed by the flavor-current two-point function coefficient~$C_F$ via~$2S_d \left(c_{TFF}+e_{TFF}\right)=dC_F$ (see~\cite{Osborn:1993cr,Erdmenger:1996yc} for details). As shown in~\cite{Hofman:2016awc, Hartman:2016dxc}, the ratio~$c_{TJJ} / C_F$ satisfies the following~$d$-dimensional collider bounds, 
\begin{equation}
\frac{(d-2)\Gamma (\half d+1)}{2(d-1)\pi ^\frac{d}{2}}\leq \frac{c _{TFF}}{C_F}\leq \frac{\Gamma (\half d +1)}{2\pi ^\frac{d}{2}}~,
\end{equation}
where the first inequality is saturated for free scalars and the second one for free fermions. In~$d = 6$ we can rewrite these inequalities as 
\begin{equation}\label{JJTinequalities}
0\leq \frac{e_{TFF}}{c_{TFF}}\leq \frac{1}{4}~,
\end{equation}
where the order of the inequalities has been reversed, i.e.~the first one is saturated for free fermions and the second one for free scalars.  

We can now apply~\eqref{JJTinequalities} to the~$\langle T J^R J^R\rangle$ correlator in SCFTs. Substituting the free-field formulas for~$c_{TRR}, e_{TRR}$ in terms of~$c_1, c_3$ in~\eqref{TRR} into these inequalities, we find that
\begin{equation}\label{c1c3ratiob}
\frac{5}{9}\leq \frac{c_1}{c_3}\leq \frac{10}{9} \qquad \Longleftrightarrow \qquad  \frac{1}{9}\leq \frac{c_2}{c_3}\leq \frac{11}{9}~.
\end{equation}
The lower bound coincides with that in~\eqref{c1c3ratio}, but the upper bound is stronger; it is saturated by a free $\CN =(1,0)$ tensor multiplet.  In terms of the free-field parameterization of the~$c_i$ conformal anomalies in~\eqref{c123nhnt}, the inequalities~\eqref{JJTinequalities} reduce to the intuitive requirement that 
\begin{equation}
n_H\geq 0~, \qquad n_T\geq 0~.
\end{equation}

\section{Anomaly Multiplet Relations on the Tensor Branch}\label{sec:dilaton}

In this section we explore how the anomaly multiplet relations derived above are reflected on the tensor branch of~$\CN=(1,0)$ SCFTs in six dimensions. This leads to a simple, intuitive argument for some of these relations.  

\subsection{Dilaton Effective Action on the Tensor Branch}

Here we closely follow the discussion in~\cite{Cordova:2015fha}, but also make some new observations. On the tensor branch, the $SU(2)_R$ symmetry is unbroken and the massless dilaton field $\varphi$ (i.e.~the Nambu-Goldstone boson associated with spontaneous conformal symmetry breaking) resides in a tensor multiplet.  The constraints of conformal symmetry on the effective action of~$\varphi$ with and without background gravity fields were analyzed in~\cite{Elvang:2012st}. The additional terms that arise in the presence of background flavor gauge fields were described in~\cite{Osborn:2015rna}. When all background gauge fields are set to zero and the background metric is taken to be the flat Minkowski metric, the minimal low-energy effective action for $\varphi$ schematically takes the form
\begin{equation}\label{6ddilgen}
\SL_\text{dilaton}=\half (\partial \varphi)^2 -b\frac{(\partial \varphi)^4}{\varphi ^3} +\Delta a\frac{(\partial \varphi)^6}{\varphi ^6}+{\cal O}(\partial ^8)~.
\end{equation}
Here, $\Delta a=a_{\text{UV}}-a_{\text{IR}}$ is the change in the conformal $a$-anomaly along the RG flow from the SCFT at the origin to the low-energy effective theory on the tensor branch.  A basic fact that will play an important role below is that the coefficient $b$ in \eqref{6ddilgen} is subject to a dispersion relation that implies its positivity \cite{Adams:2006sv},
\begin{equation}\label{binequality}
b\geq 0~.
\end{equation}
This inequality is saturated if and only if~$\varphi$ is a free field with trivial scattering S-matrix.  

The constraints of $\CN =(1,0)$ supersymmetry on the dilaton effective action were analyzed in~\cite{Cordova:2015fha}. In the absence of non-trivial background fields, the~$\CN=(1,0)$ dilaton effective action on the tensor branch is given by the supersymmetrization of~\eqref{6ddilgen}, with~$\varphi$ residing in a tensor multiplet together with its superpartners~$\psi^i_\alpha$ and~$B$. Here~$\psi^i_\alpha$ is a Majorana-Weyl fermion, and~$B$ is a two-form gauge field, whose field strength~$H = dB$ is self dual (i.e.~$H = * H$). The upshot of this analysis is two-fold~\cite{Cordova:2015fha}: the first conclusion is that the supersymmetric completion of the leading 4-derivative term in~\eqref{6ddilgen} is of the form
 \begin{equation}\label{onezerodil}
-b\frac{(\partial \varphi)^4}{\varphi ^3} \qquad \longrightarrow \qquad  -\frac{b}{\langle \varphi \rangle^3}Q^8 (\delta\varphi)^4 + \cdots~,
\end{equation}
where the ellipisis denotes higher-order terms in the expansion of~$\varphi = \langle \varphi\rangle + \delta\varphi$ in fluctuations around its vev. The second conclusion is that the 6-derivative term in~\eqref{6ddilgen} proportional to~$\Delta a$ is in fact related to the term in~\eqref{onezerodil} by supersymmetry, which implies the following universal quadratic relation,
\begin{equation}
\Delta a = \frac{98304\pi ^3}{7}b^2~.
\end{equation}
This immediately implies positivity of~$\Delta a$, and hence the~$a$-theorem, for this class of flows~\cite{Cordova:2015fha}.

In the presence of background fields, the dilaton effective action~\eqref{6ddilgen} gets extended in various ways. Among these extensions there are certain additional terms that are needed to compensate any apparent mismatch between the 't Hooft anomalies of the SCFT at the origin and the low-energy theory on the tensor branch. These terms are Green-Schwarz (GS) like couplings that involve the dynamical two-form gauge field $B$ residing in the tensor multiplet and various background field strengths or background curvatures~\cite{Ohmori:2014kda,Intriligator:2014eaa}, and their supersymmetric completions. Given, for instance, a background flavor field strength~$F_{\mu \nu}$, the corresponding GS term takes the form
\begin{equation}\label{flavorbackgrounds}
\SL _{GS} \supset n_F B\wedge c_2(F)~. 
\end{equation}
It follows that $c_2(F)$ (the background flavor instanton density) acts as a source for the dynamical~$B$-field, whose field-strength is~$H$,\begin{equation}
dH\supset n_F c_2(F)~.
\end{equation}
This in turn implies that~$n_F$ is an integer. The supersymmetric completion of~\eqref{flavorbackgrounds} contains a dilaton couplling~$\sim n_f \varphi \tr \left(F_{\mu\nu} F^{\mu\nu}\right)$. Together with~\eqref{flavorbackgrounds}, this Lagrangian is similar to the interacting gauge-tensor Lagrangians used to described~$\CN=(1,0)$ SCFTs in~\cite{Seiberg:1996qx}, except that here the gauge fields are fixed backgrounds associated with global flavor symmetries while only the tensor multiplet is dynamics. 

In addition to the flavor background fields discussed above, the tensor multiplet containing~$B$ and~$\varphi$ also couples to supergravity background fields -- in particular a background metric and background gauge fields for the~$SU(2)_R$ symmetry. Here we will briefly review these terms, following~\cite{Cordova:2015fha}. The GS terms involving the dynamical~$B$-field and supergravity background fields are  
\begin{equation}\label{GSxy}
\SL_{GS}\supset  B\wedge \left(xc_{2}(R)+yp_{1}(T)\right)~,
\end{equation}
where $x,y$ are real coefficients. These terms account for the following mismatches~$\Delta \alpha = \alpha_\text{UV} - \alpha_\text{IR}$ etc.~in the~$SU(2)_R$ and diffeomorphism 't Hooft anomaly coefficients~\eqref{alphadelta},
\begin{equation}\label{resform}
\Delta \alpha \sim x^{2}~, \qquad \Delta \beta  \sim 2 x y~, \qquad\Delta \gamma \sim y^{2}~, \qquad\Delta \delta =0~.
\end{equation}
It was shown in~\cite{Cordova:2015fha} that the GS terms~\eqref{GSxy} are related to certain~$R^2$ supergravity terms~\cite{Bergshoeff:1986wc},
\begin{equation}
\SL_{R^{2}}\sim \langle \varphi\rangle \sqrt{g}\left(\left(y-\frac{x}{4}\right)R^{\mu\nu\rho\sigma}R_{\mu\nu\rho\sigma}+\frac{3}{2}xR_{[\mu\nu}^{\phantom{\mu\nu}\mu\nu}R_{\rho\sigma]}^{\phantom{\rho\sigma}\rho\sigma}\right)~.
\end{equation}
Specializing to a conformally flat background and using results from~\cite{Elvang:2012st}, this shows that the leading four-dilaton interaction~\eqref{6ddilgen} is determined by the GS coefficeints~$x$ and~$y$,
\begin{equation}
\SL_{R^{2}}\qquad  \longrightarrow \qquad - b \, \frac{(\partial \varphi)^{4}}{\varphi^{3}}~, \qquad  b\sim y-x~. 
\end{equation}
The inequality \eqref{binequality} then shows that~$y \geq x$. This inequality can only be saturated if the dilaton is a free tensor multiplet, in which case there is no RG flow to begin with. In particular, $\Delta a \sim b^2 \sim (y-x)^2$ vanishes in this case~\cite{Cordova:2015fha}.

\subsection{The~$c_i$ Conformal Anomalies on the Tensor Branch}

We will now use the tensor branch to argue for the anomaly multiplet formulas~\eqref{csare} that determine the~$c_i$ conformal anomalies in terms of the 't Hooft anomalies~$\alpha, \beta, \gamma, \delta$. We will not provide a complete analysis of these anomaly multiplet relations on the tensor branch, which would require the dilaton effective action coupled to background supergravity fields up to six-derivative order. Instead, we will make the plausible assumption that all six-derivative terms responsible for anomaly matching in the dilaton effective action arise from the completion of four-derivative GS terms and their superpartners. (This assumption was explicitly demonstrated in~\cite{Cordova:2015fha} for the six-derivative terms associated with the~$a$ conformal anomaly.) We also postulate that the~$c_i$ anomalies obey linear anomaly multiplet relations of the form 
\begin{equation}\label{pqrs}
c_{i}=p_{i}\alpha +q_{i}\beta +r_{i}\gamma+s_{i}\delta
\end{equation}
Using~\eqref{resform}, it follows that the change in these anomalies under the RG flow onto the tensor branch is
\begin{equation}\label{eq:deltaci}
\Delta c_{i} \sim p_{i}x^{2} +2 q_{i}xy +r_{i}y^{2}~.
\end{equation}

At the special locus $x=y$, the dilaton is free and there is no non-trivial RG flow, so that all $\Delta c_i$ must vanish. It follows that the quadratic polynomials in~\eqref{eq:deltaci} must factor as~$\Delta c_i = (y-x)(u_ix+v_iy)$ (recall that this is the case for~$\Delta a \sim (y-x)^2$), so that 
\begin{equation}\label{linrelqpr}
p_{i}+2{q}_{i}+r_{i}=0~.
\end{equation}
We can use these linear relations to reduce the number of unknown coefficients in~\eqref{pqrs} from four to three for each~$c_i$. Ideally, these coefficients should be derived by constructing the full anomaly multiplet, but we can also determine them by comparing to the free~$\CN=(1,0)$ hyper- and tensor multiplets, as well as to the interacting~$\CN=(2,0)$ theories. As was explained in section~\ref{sec:introam}, this set of reliable unitary examples allows us to fix three of the four coefficients in~\eqref{pqrs}. Together with~\eqref{linrelqpr}, this allows us to fix the anomaly multiplet relations for the~$c_i$ conformal anomalies in~\eqref{csare} while avoiding the non-unitary examples considered in~\cite{Yankielowicz:2017xkf, Beccaria:2017dmw}.

\subsection{Flavor Conformal Anomalies on the Tensor Branch} 

In four dimensional SCFTs, the flavor-current two-point function  coefficient~$\tau_F$ is determined by the 't Hooft triangle anomaly of two flavor currents and one~$R$-current. This follows from the~$d = 4$ multiplet of anomalies in the presence of background supergravity and flavor gauge fields ~\cite{Anselmi:1997am}. We anticipate that
a similar relation exists for six-dimensional SCFTs, which should relate the flavor-current two-point function coefficient $\tau _{F}$ to 't Hooft anomalies involving background supergravity and flavor gauge fields. Since~$\tau _{F}$ scales quadratically with the charges (e.g. in free field theory), such a putative relation must involve a mixed 't Hooft anomaly with two flavor background gauge fields. The remaining two background fields can either be~$SU(2)_R$ or gravity backgrounds, i.e.~we consider mixed flavor-$SU(2)_R$ and flavor-diffeomorphism anomalies. We therefore postulate a linear relation between $\tau  _{F}$ and the 't Hooft anomaly coefficients $\alpha _{F^2R^2}$ and $\alpha _{F^2T^2}$ in the anomaly 8-form polynomial,
\begin{equation}\label{eq:ffmixedanom}
 \CI _8\supset \frac{1}{4!}\left(\alpha _{F^2R^2}c_2(F)c_2(R)+\alpha _{F^2T^2}c_2(F)p_1(T)\right)~.
\end{equation}

Recall from the discussion above~\eqref{tau1norm} that we normalize~$\tau_F$ so that a free~$\CN =(2,0)$ tensor multiplet contributes has~$\tau_2^{Sp(4)_R} = 1$. In this normalization, the~$\tau_2$ conformal anomaly and the 't Hooft anomalies in~\eqref{eq:ffmixedanom} for some known examples are summarized in Table \ref{knownFA}.
 \begin{table}[h]
\centering
\begin{tabular}{!{\VRule[1pt]}c!{\VRule[1pt]}c!{\VRule[1pt]}c!{\VRule[1pt]}c!{\VRule[1pt]} }
\specialrule{1.2pt}{0pt}{0pt}
{\bf Theory} & $\bf \tau _F$ &  $\bf \alpha _{F^2R^2}$ & $\bf \alpha _{F^2T^2}$  \\
\specialrule{1.2pt}{0pt}{0pt}
\multirow{2}{*}{ $(1,0)$ hypermultiplet in flavor representation $r$ }& \multirow{2}{*}{$2T_2(r)$} &  \multirow{2}{*}{$0$} & \multirow{2}{*}{$T_2(r)$}  \\
 &  & &   \\
\hline
 \multirow{2}{*}{$(2,0)$ theory with ADE algebra $\frak{g}$}& \multirow{2}{*}{$4h^\vee _{\frak{g}}d_{\frak{g}}+r_{\frak{g}}$} &  \multirow{2}{*}{$-2h^\vee _{\frak{g}}d_{\frak{g}} $} & \multirow{2}{*}{$\half r_{\frak{g}}$} \\
 &  & &   \\
\specialrule{1.2pt}{0pt}{0pt}
 \end{tabular}
\caption{Some conformal and 't Hooft anomalies related to flavor symmetries. For the~$(2,0)$ theories, the flavor symmetry is~$SU(2)_F \subset Sp(4)_R$, which is related to~$SU(2)_R \subset Sp(4)_R$ by a Weyl reflection, whence~$\tau_F = \tau_1^R = c_3$ (see the discussion below~\eqref{susytaurrel}).  }
\label{knownFA}
\end{table}
The examples in Table~\ref{knownFA} are sufficient to determine the proposed linear relation between~$\tau _{F}$ and $\alpha _{F^2R^2}$, $\alpha _{F^2T^2}$,
\begin{equation}\label{taualphaagain}
\tau _{F}=2\alpha _{F^2T^2}-2\alpha _{F^2R^2}~.
\end{equation}
This is the relation in~\eqref{taualpha}.

We can now subject the proposed formula~\eqref{taualphaagain} to a stringent consistency check by considering an RG flow onto the tensor branch, under which 
\begin{equation}\label{taumatch}
\Delta \tau _{F}=2\Delta \alpha _{F^2T^2}-2\Delta \alpha _{F^2R^2}~.
\end{equation} 
The anomaly mismatches~$\Delta \alpha_{F^2 T^2}$ and~$\Delta \alpha_{F^2 R^2}$ are accounted for by the GS terms for supergravity background fields in~\eqref{GSxy}, in conjunction with the flavor GS term~\eqref{flavorbackgrounds},
\begin{equation}\label{GSxyn}
\SL _{GS}\sim B\wedge \left(x c_2(R)+yp_1(T)+n_F c_2(F)\right)~.
\end{equation}
This leads to 't Hooft anomaly matching contributions beyond~\eqref{resform},
\begin{equation}
\Delta \alpha _{F^2T^2} \sim n_F y~, \qquad \Delta \alpha _{F^2 R^2} \sim n_F x~, \qquad \Delta \alpha _{F^2F^2} \sim  n_F^2~,
\end{equation}
with the same overall proportionality factor as in~\eqref{resform}. Substituting into~\eqref{taumatch}, we find that 
\begin{equation}\label{deltataunyx}
\Delta \tau _{F}=2(\Delta \alpha _{F^2T^2}-\Delta \alpha _{F^2R^2})\sim 2 n_F\left(y-x\right)~.
\end{equation}
This is indeed proportional to $b\sim y-x$, and trivializes as expected when~$x = y$.

\section{Examples}\label{sec:examples}

\subsection{$\CN =(2,0)$ Theories}

These theories have an~$Sp(4)_R$ symmetry. Thinking of them in~$\CN=(1,0)$ language, it is natural to focus on its~$SU(2)_F \times SU(2)_R$ subgroup, with~$SU(2)_F$ a flavor symmetry. In our normalization, the conformal anomalies of the interacting~$\CN=(2,0)$ theory based on an ADE lie algebra~$\frak{g}$ are given by
\begin{equation}
a_{\frak{g}}=\frac{16}{7} h_{\frak{g}}^\vee d_{\frak{g}}+r_{\frak{g}}~, \qquad c_{\frak{g},i}=\tau _{F}=\tau _{1}^R=4h_{\frak{g}}^\vee d_{\frak{g}}+r_{\frak{g}}~.
\end{equation}

Consider a rank-one tensor branch of the type~$\frak g$ theory, associated with the adjoint breaking pattern $\frak{g}\to \frak{h}+\frak{u}(1)$.  The apparent change in the anomaly polynomial is~\cite{Intriligator:2000eq}
\begin{equation}
4!\Delta {\cal I}_8=\Delta k \, p_2(Sp(4)_R)\;\longrightarrow\; \Delta k \left(c_2(L)-c_2(R)\right)^2~,  \; \Delta k\equiv k(\frak{g})-k(\frak{h})~, \; k(\frak{g})\equiv h^
\vee _{\frak{g}}d_{\frak{g}}~. 
\end{equation}
This requires GS anomaly matching terms~\eqref{GSxyn}, with~$y=0$ and $x=n_F \sim -\sqrt{\Delta k/6}$.\footnote{~The anomaly matching of the full~$Sp(4)_R$ symmetry is more involved and requires the Hopf-Wess-Zumino term of~\cite{Intriligator:2000eq}.} The simplest example in this class occurs for $\frak{g}=\frak{a}_1$ and $\frak{h}$ trivial, so that $\Delta k/6=1$.  The interacting~$\frak a_1$ SCFT at the origin has $c_i  = \tau _{F}=\tau_1^R = 25$ and $a= {103 \over 7}$. This theory was explored using numerical bootstrap techniques in~\cite{Beem:2015aoa}.

\subsection{Small $E_8$ Instanton SCFTs}

We can apply our general formulas to determine the conformal anomalies of the~$\CN=(1,0)$ SCFTs ${\cal E}[N]$ that describe~$N$ small $E_8$ instantons in string theory, or alternatively~$N$ M5 branes probing an end-of-the-world M9 brane in M-theory. These theories have an~$SU(2)_F \times E_8$ flavor symmetry. Aspects of the corresponding flavor-current correlators were discussed in~\cite{Cheung:1997id, Chang:2017xmr}. Note that our definition of the~${\cal E}[N]$ theory includes the free, decoupled hypermultiplet describing overall translations of the~$N$ M5 branes in the four transverse directions. 

The anomaly polynomial of the ${\cal E}[N]$ SCFT was found in~\cite{Ohmori:2014pca},
\begin{equation}\label{E8anomalies}
\CI _{\CE [N]}=\frac{N^3}{6} \chi _4^2 +\half N^2 \chi _4I_4+N(\half I_4^2-I_8)~.
\end{equation}
where $\chi _4\equiv c_2(F)-c_2(R)$ and 
\begin{equation}
I_4\equiv \frac{1}{4}(-2 c_2(R)-2c_2(F)+p_1(T))+c_2(E_8)~,
\end{equation}
\begin{equation}
I_8\equiv \frac{1}{48}\left(\chi _4^2 +p_2(T)-\frac{1}{4}(-2c_2(R)-2c_2(F)-p_1(T))^2\right)~.
\end{equation}
As mentioned in~\cite{Ohmori:2014pca}, the anomaly polynomial~\eqref{E8anomalies} exhibits an interesting behavior under the combined transformation $N\to -N$ and $SU(2)_F\leftrightarrow SU(2)_R$, which takes $\CI _{\CE [N]}\to -\CI _{\CE [N]}$.  This was interpreted as exchanging branes with anti-branes, and $\CN =(1,0)$ with $\CN =(0,1)$. We will choose~$N > 0$, so that $SU(2)_R$ is the~$\CN=(1,0)$ R-symmetry, while $SU(2)_F$ is a flavor symmetry. 

Comparing with~\eqref{alphadelta}, the anomaly polynomial~\eqref{E8anomalies} gives
\begin{equation}
\alpha= N(4N^{2}+6N+3)~, \qquad  \beta= -\frac{N}{2}(6N+5)~,\qquad \gamma= \frac{7N}{8}~, \qquad \delta=-\frac{N}{2} ~.
\end{equation}
Substituting into~\eqref{ais} and~\eqref{csare} leads to the following~$a$ and~$c_i$ conformal anomalies, 
\begin{eqnarray}\label{Eanoms}
a=\frac{64}{7}N^{3}+\frac{144}{7}N^{2}+\frac{99}{7}N~, \hspace{.5in}c_1=16 N^3+38N^2+27N~,\\
c_2=16 N^3 + 34 N^2+21N~, \hspace{.5in}c_3=16 N^3+42 N^2+33N~. \nonumber
\end{eqnarray}
To $\CO(N^3)$, these anomalies agree with those obtained from a naive~$\Z _2$ orbifold of the $\frak{a}_{N-1}$ $\CN =(2,0)$ theory, as expected from the M5 brane construction of the~${\cal E}[N]$ theory. Note however that the~$c_i$ anomalies in~\eqref{Eanoms} already differ at $\CO(N^2)$.  

As was already stated above, the ${\cal E}[N]$ SCFTs have an $SU(2)_F\times E_8$ flavor symmetry. The associated 't Hooft anomalies can be extracted from the anomaly polynomial~\eqref{E8anomalies}, e.g. 
\begin{equation}
\alpha _{F^2T^2}=3N^2 - \frac{5}{2}N~, \qquad \alpha _{F^2 R^2}=-8N^3+8N~.
\end{equation}
Substituting into~\eqref{taualpha} gives the $SU(2)_F$ two-point function coefficient,
\begin{equation}\label{su2L}
\tau _{F}=16 N^3+6N^2-21N~.
\end{equation}
For comparison, the $SU(2)_R$ current two-point function coefficient is given by (see~\eqref{susytaurrel})
\begin{equation}\label{su2R}
\tau _{1}^R=c_3=16 N^3+42N^2+22N~.
\end{equation}
The fact that $\tau_F$ and $\tau_1^R$ coincide at leading $\CO(N^3)$ is again consistent with a naive~$\Z_2$ orbifold of the~$\frak a_{N-1}$~$\CN=(2,0)$ theory, and again the two expressions differ at~$\CO(N^2)$. Note that~\eqref{su2L} and~\eqref{su2R} are not simply related by the $N\to -N$ transformation mentioned above. This is not a contradiction, since it need not be the case that replacing $N\to -N$ exchanges~$SU(2)_F$ and~$SU(2)_R$ for all purposes.\footnote{~An analogous phenomenon occurs in certain~$d = 4$, ${\cal N}=2$ SCFTs~\cite{Aharony:2007dj} with similar brane realizations.}    As a check of~\eqref{su2L}, note that when~$N=1$ the~$SU(2)_F$ flavor symmetry only acts on the free, decoupled hypermultiplet associated with M5 brane motion in the transverse directions. Indeed substituting~$N=1$ in~\eqref{su2L} gives $\tau _{F}=1$,  as expected for a free hypermultiplet.  

The anomaly coefficients in~\eqref{E8anomalies} that involve the $E_8$ background gauge field are
\begin{equation}
\alpha _{E_8^2 R^2}=-12(N^2+N)~, \qquad \alpha _{E_8^2T^2}=6N~.
\end{equation}
Substituting into~\eqref{taualpha} then gives the~$E_8$ two-point function coefficient,
\begin{equation}\label{eq:taueeight}
\tau _{E_8}=12(2N^2+3N)~.
\end{equation}
This result has already appeared in~\cite{Chang:2017xmr}.

Finally, let us consider a tensor-branch deformation corresponding to moving one M5 brane away from the M9 brane and the other~$N-1$ M5 branes, so that~${\cal E}[N] \rightarrow {\cal E}[N-1] + {\cal E}[1]$. 
The anomaly mismatch is accounted for by a GS mechanism~\cite{Intriligator:2014eaa},
\begin{equation}
\Delta \CI _8=\half X_4^2~,\qquad  X_4=(N-1)c_2(F)-Nc_2(R)+\frac{1}{4}p_1(T)+c_2(E_8)~.
\end{equation}
Since the GS term~\eqref{GSxy} is proportional to~$B \wedge \CI_4$, we conclude that 
\begin{equation}\label{eq:xysmallee}
x \sim - N~, \qquad y\sim \frac{1}{4}~.
\end{equation}
Using~\eqref{su2L}, \eqref{su2R}, and~\eqref{eq:taueeight}, we can evaluate
\begin{equation}\label{eq:deltasinsee}
\Delta \tau _1^R=\Delta c_3 = 48(N+\half)(N+\frac{1}{4})~, \; \Delta \tau _F=48(N-1)(N+\frac{1}{4})~, \; \Delta \tau _{E_8}=48(N+\frac{1}{4})~.
\end{equation}
Since it follows from~\eqref{eq:xysmallee} that $N+\frac{1}{4} \sim y-x$, it is indeed true that all quantities in~\eqref{eq:deltasinsee} are proportional to $y-x$, as we argued on general grounds.

\subsection{Holographic Examples}

Consider $d = 6$ SCFTs with AdS$_7$ holographic duals; see e.g.\ \cite{DeLuca:2018zbi} and references therein for examples.  The  AdS$_{d+1}/$CFT$_d$ dictionary relates the flavor-current two-point function coefficient $C_F$ to $L^{d-3}g_{FF}^{-2}$ ~\cite{Freedman:1998tz},
where $L$ is the AdS$_{d+1}$ length scale, and $g_{FF}^{-2}$ is the coefficient of the bulk Yang-Mills term corresponding to the flavor symmetry in the boundary CFT, $S_\text{bulk}\supset \int (-\frac{1}{4}g_{FF}^{-2}\Tr F_F\wedge *F_F)$.    In our conventions and~$d = 6$ we have
\begin{equation}
\tau_1^F=\frac{5}{3} \, 2^5 L^3 g_{FF}^{-2}~.
\end{equation}
The fact that $\tau_2^F=\rho^F=0$ in SCFTs should follow from an analysis of AdS$_7$ supergravity. Moreover, the relation~\eqref{taualpha} implies that the $g_{FF}^{-2}$ Yang-Mills coefficient of the AdS$_7$ supergravity theory is related to the bulk Chern-Simons terms responsible for the~$\alpha _{F^2T^2}$ and $\alpha _{F^2R^2}$ 't Hooft anomalies on the boundary. See~\cite{Gherghetta:2002nq} for some related comments.   

\bigskip

\section*{Acknowledgements}

\noindent We are grateful to C.-M.~Chang, Y.-H.~Lin, and A.~Stergiou for helpful discussions. TD is supported by a DOE Early Career Award under DE-SC0020421 and by the Mani L. Bhaumik Presidential Chair in Theoretical Physics at UCLA. KI is supported by DOE grant DE-SC0009919, the Dan Broida Chair, and by Simons Investigator Award 568420.

\bigskip

\bibliographystyle{utphys}
\bibliography{references}

\end{document}